\documentstyle[preprint,aps,epsf,psfig,eqsecnum]{revtex}
\tightenlines
\begin{document}
\def\be{\begin{eqnarray}}
\def\en{\end{eqnarray}}
\def\non{\nonumber}
\def\la{\langle}
\def\ra{\rangle}
\def\ep{\varepsilon}
\def\hep{\hat{\varepsilon}}
\def\ek{{\vec{\ep}_\perp\cdot\vec{k}_\perp}}
\def\epp{{\vec{\ep}_\perp\cdot\vec{P}_\perp}}
\def\kp{{\vec{k}_\perp\cdot\vec{P}_\perp}}
\def\lsim{ {\ \lower-1.2pt\vbox{\hbox{\rlap{$<$}\lower5pt\vbox{\hbox{$\sim$}
}}}\ } }
\def\gsim{ {\ \lower-1.2pt\vbox{\hbox{\rlap{$>$}\lower5pt\vbox{\hbox{$\sim$}
}}}\ } }
\def\dk{\partial\!\cdot\!K}
\def\pr{{\sl Phys. Rev.}~}
\def\prl{{\sl Phys. Rev. Lett.}~}
\def\pl{{\sl Phys. Lett.}~}
\def\np{{\sl Nucl. Phys.}~}
\def\zp{{\sl Z. Phys.}~}

\draft
\title{
\begin{flushright}
{\normalsize ~~~IP-ASTP-04-96
}
\end{flushright}
\Large\bf
Mesonic Form Factors and the Isgur-Wise Function on the Light-Front
}
\author{{\bf Hai-Yang Cheng$^{a}$},
{\bf Chi-Yee Cheung$^{a}$},
{\bf Chien-Wen Hwang$^{a,b}$}
}
\vskip 1.0cm
\address{$^{a}$Institute of Physics, Academia Sinica, Taipei, Taiwan 115, 
Republic of China}
\address{$^{b}$Department of Physics, National Taiwan University, Taipei,
Taiwan 107, Republic of China}
%
\date{July, 1996}
\maketitle
\begin{abstract}
   Within the light-front framework, form factors for $P\to P$ and $P\to V$
transitions ($P$: pseudoscalar meson, $V$: vector meson) due to the
valence-quark configuration are calculated directly in the entire physical
range of momentum transfer. The behavior of the form factors in the infinite
quark mass limit are examined to see if the requirements of heavy-quark 
symmetry are fulfilled. We find that the Bauer-Stech-Wirbel type of 
light-front wave function fails to give a correct normalization for the 
Isgur-Wise function at 
zero recoil in $P\to V$ transition. Some of the 
$P\to V$ form factors are found to depend on the recoiling direction of the 
daughter mesons relative to their parents. Thus, the inclusion of the 
non-valence contribution arising from 
quark-pair creation is mandatory in
order to ensure that the physical form factors are independent of the 
recoiling direction. The main feature of the non-valence contribution is 
discussed.

\vskip 0.25 true cm
PACS numbers: 13.20, 14.40.J
\end{abstract}
\newpage
\baselineskip .29in

\section{Introduction}
   The hadronic matrix element of weak $P\to P$ transition ($P$: pseudoscalar
meson) is described by two form factors, whereas in general it requires
four form factors to parametrize the weak matrix element for $P\to V$
transition ($V$: vector meson). Heavy quark symmetry predicts that, all
the mesonic form factors in the infinite quark mass limit $m_Q\to\infty$
are related to a single universal Isgur-Wise function \cite{iw}. The symmerty 
breaking $1/m_Q$ 
corrections can be studied in a systematic framework,
namely the heavy quark effective theory (for a review, see \cite{Neu94}).
The Isgur-Wise function is normalized to unity at zero recoil, but otherwise 
remains unknown. Phenomenologically, the hadronic form factors can be 
evaluated in various models among which the quark model is a popular one. 
However, since usual 
quark-model wave functions best resemble meson states in the rest frame or 
where the meson velocities are small, hence the form
factors calculated in non-relativistic quark model or the MIT bag model are 
trustworthy only when the recoil momentum of the daughter meson relative to 
the parent meson is small.

   As the recoil momentum increases (corresponding to a decreasing $q^2$),
we have to start considering relativistic effects
seriously. In particular, at the maximum recoil point $q^2=0$ where the final
meson could be highly relativistic, there is no reason to expect that
the non-relativistic quark model is still applicable. A consistent treatment
of the relativistic effects of the quark motion and spin in a bound state
is a main issue of the relativistic quark model. To our knowledge,
the light-front quark model \cite{Ter,Chung} is the only relativistic quark
model in which a consistent and fully relativistic treatment of quark spins
and the center-of-mass motion can be carried out. This model has many
advantages. For example, the light-front wave function is manifestly Lorentz
invariant as it is expressed in terms of the momentum fraction variables
(in ``+" components) in analog to the parton distributions in the infinite
momentum frame. Moreover, hadron spin can also be correctly constructed using 
the so-called Melosh rotation. The kinematic subgroup of the light-front 
formalism has the maximum number of interaction-free generators including the 
boost operator which describes the center-of-mass motion of the bound state
(for a review of the light-front dynamics and light-front QCD, see 
\cite{Zhang}).

   The light-front quark model has been applied in the past to study the
heavy-to-heavy and heavy-to-light weak decays form factors 
\cite{Jaus,Don94a,Don94b,Don95}. However, the weak form factors were 
calculated only for $q^2\leq 0$, whereas physical decays occur in the
time-like region $0\leq q^2\leq (M_i-M_f)^2$, with $M_{i,f}$ being the 
initial and final meson masses. Hence extra assumptions are needed to 
extrapolate the form factors to cover the entire 
range of momentum transfer. In \cite{Jaus96}
an ansatz for the $q^2$ dependence was made to extrapolate the form factors 
in the space-like 
region to the time-like region. Based on the dispersion
formulation, form factors at $q^2>0$ were obtained in \cite{Mel}
by performing an analytic continuation from the space-like $q^2$ region. 
Finally, the weak form factors for $P\to P$ transition were calculated
in \cite{Cheung2,Sima,Simb} for the first time for the entire range of
$q^2$, so that additional extrapolation assumptions are no
longer required. This is based on the observation \cite{Dubin} that in 
the frame where the momentum transfer is purely longitudinal, i.e., 
$q_\perp=0$, $q^2=q^+q^-$ covers the entire range of momentum transfer.
The price one has to pay is that, besides the
conventional valence-quark contribution, one must also consider the 
non-valence configuration
(or the so-called $Z$-graph) arising from quark-pair creation from
the vacuum (see Fig.~1). The non-valence contribution vanishes if $q^+=0$,
but is supposed to be important for heavy-to-light transition near zero
recoil \cite{Jaus,Jaus96,Dubin,Saw}. Unfortunately, a reliable way of 
estimating the $Z$-graph contribution is still lacking.

    In the present paper we calculate the $P\to V$ form factors 
directly at time-like momentum transfers for the first time. We then
study the mesonic form factors in the infinite quark mass limit to check
if the light-front model calculations respect heavy quark symmetry. We are
able to compute the Isgur-Wise function exactly since the non-valence
contribution vanishes in the heavy-quark limit. It turns out that not all
light-front wave functions give a correct normalization
for the Isgur-Wise function at zero recoil in $P\to V$ decay. In other words,
the requirement of heavy quark symmetry can be utilized to rule out
certain phenomenological wave functions.

    Another issue we would like to address in this work has to do with 
reference frame dependence of form factor. For a given $q^2$, one can choose 
whether the recoiling 
daughter meson moves in the positive or negative $z$-direction relative to
the parent meson, which we call the ``+" and ``$-$" reference frame,
respectively. For some form factors in $P\to V$ transition, namely
$A_0,~A_1,~V$, valence-quark and non-valence contributions are
separately dependent on the choice of the ``+" or ``$-$" frame, but their 
sum should not. This demonstrates the fact that it is mandatory to take into 
account the non-valence configuration in order to have physical 
predictions for form factors. This issue will be discussed in more details 
in Sections ~IID and IVC.

  This paper is organized as follows.  In Section II, the basic
theoretical formalism is given and form factors for $P\to P$ and $P\to V$
transitions are derived. Section III is devoted to the discussion of the 
Isgur-Wise function. 
Numerical results are present and discussed in Section
IV, and finally a summary is given in Section V.

\section{Framework}
    We will describe in this section the light-front approach for the
calculation of the weak mesonic form factors for pseudoscalar-to-pseudoscalar 
and pseudoscalar-to-vector transitions. The hadronic matrix elements will be 
evaluated at time-like momentum transfers, namely the physically accessible
kinematic region $0\leq q^2\leq q^2_{\rm max}$. 

A meson bound state consisting of a quark $q_1$ and
an antiquark $\bar q_2$ with total momentum $P$
and spin $S$ can be written as
\begin{eqnarray}
        |M(P, S, S_z)\rangle
                =\int &&\{d^3p_1\}\{d^3p_2\} ~2(2\pi)^3 \delta^3(\tilde
                P-\tilde p_1-\tilde p_2)~\nonumber\\
        &&\times \sum_{\lambda_1,\lambda_2}
                \Psi^{SS_z}(\tilde p_1,\tilde p_2,\lambda_1,\lambda_2)~
                |q_1(p_1,\lambda_1) \bar q_2(p_2,\lambda_2)\rangle,
\end{eqnarray}
where $p_1$ and $p_2$ are the on-mass-shell light-front momenta,
\begin{equation}
        \tilde p=(p^+, p_\bot)~, \quad p_\bot = (p^1, p^2)~,
                \quad p^- = {m^2+p_\bot^2\over p^+},
\end{equation}
and
\begin{eqnarray}
        &&\{d^3p\} \equiv {dp^+d^2p_\bot\over 2(2\pi)^3}, \nonumber \\
        &&|q(p_1,\lambda_1)\bar q(p_2,\lambda_2)\rangle
        = b^\dagger_{\lambda_1}(p_1)d^\dagger_{\lambda_2}(p_2)|0\rangle,\\
        &&\{b_{\lambda'}(p'),b_{\lambda}^\dagger(p)\} =
        \{d_{\lambda'}(p'),d_{\lambda}^\dagger(p)\} =
        2(2\pi)^3~\delta^3(\tilde p'-\tilde p)~\delta_{\lambda'\lambda}.
                \nonumber
\end{eqnarray}
In terms of the light-front relative momentum
variables $(x, k_\bot)$ defined by
\begin{eqnarray}
        && p^+_1=x_1 P^+, \quad p^+_2=x_2 P^+, \quad x_1+x_2=1, \nonumber \\
        && p_{1\bot}=x_1 P_\bot+k_\bot, \quad p_{2\bot}=x_2
                P_\bot-k_\bot,
\end{eqnarray}
the momentum-space wave-function $\Psi^{SS_z}$
can be expressed as
\begin{equation}
        \Psi^{SS_z}(\tilde p_1,\tilde p_2,\lambda_1,\lambda_2)
                = R^{SS_z}_{\lambda_1\lambda_2}(x,k_\bot)~ \phi(x, k_\bot),
\end{equation}
where $\phi(x,k_\bot)$ describes the momentum distribution of the
constituents in the bound state, and $R^{SS_z}_{\lambda_1\lambda_2}$
constructs a state of definite spin ($S,S_z$) out of light-front
helicity ($\lambda_1,\lambda_2$) eigenstates.  Explicitly,
\begin{equation}
        R^{SS_z}_{\lambda_1 \lambda_2}(x,k_\bot)
                =\sum_{s_1,s_2} \langle \lambda_1|
                {\cal R}_M^\dagger(1-x,k_\bot, m_1)|s_1\rangle
                \langle \lambda_2|{\cal R}_M^\dagger(x,-k_\bot, m_2)
                |s_2\rangle
                \langle {1\over2}s_1
                {1\over2}s_2|SS_z\rangle,
\end{equation}
where $|s_i\rangle$ are the usual Pauli spinor,
and ${\cal R}_M$ is the Melosh transformation operator:
\begin{equation}
        {\cal R}_M (x,k_\bot,m_i) =
                {m_i+x_iM_0+i\vec \sigma\cdot\vec k_\bot \times \vec n
                \over \sqrt{(m_i+x_i M_0)^2 + k_\bot^2}},
\end{equation}
with $\vec n = (0,0,1)$, a unit vector in the $z$-direction, and
\begin{equation}
        M_0^2={ m_1^2+k_\bot^2\over x_1}+{ m_2^2+k_\bot^2\over x_2}.
\label{M0}
\end{equation}
In practice it is more convenient to use the covariant form for
$R^{SS_z}_{\lambda_1\lambda_2}$ \cite{Jaus}:
\begin{equation}
        R^{SS_z}_{\lambda_1\lambda_2}(x,k_\bot)
                ={\sqrt{p_1^+p_2^+}\over \sqrt{2} ~{\widetilde M_0}}
        ~\bar u(p_1,\lambda_1)\Gamma v(p_2,\lambda_2), \label{covariant}
\end{equation}
where
\begin{eqnarray}
        &&{\widetilde M_0} \equiv \sqrt{M_0^2-(m_1-m_2)^2}, \nonumber\\
        &&\Gamma=\gamma_5 \qquad ({\rm pseudoscalar}, S=0), \\
        &&\Gamma=-\not{\! \hep}(S_z)+{\hep\cdot(p_1-p_2)
                \over M_0+m_1+m_2} \qquad ({\rm vector}, S=1), \nonumber
\end{eqnarray}
with
\begin{eqnarray}
        &&\hep^\mu(\pm 1) =
                \left[{2\over P^+} \vec \varepsilon_\bot (\pm 1) \cdot
                \vec P_\bot,\,0,\,\vec \varepsilon_\bot (\pm 1)\right],
                \quad \vec \varepsilon_\bot
                (\pm 1)=\mp(1,\pm i)/\sqrt{2}, \nonumber\\
        &&\hep^\mu(0)={1\over M_0}\left({-M_0^2+P_\bot^2\over
                P^+},P^+,P_\bot\right).   \label{polcom}
\end{eqnarray}
Note that the longitudinal polarization 4-vector $\hep^\mu(0)$ given above 
is not exactly the same as that of the vector meson [cf. Eq.(\ref{polvec})].
We normalize the meson state as
\begin{equation}
        \langle M(P',S',S'_z)|M(P,S,S_z)\rangle = 2(2\pi)^3 P^+
        \delta^3(\tilde P'- \tilde P)\delta_{S'S}\delta_{S'_zS_z}~,
\label{wavenor}
\end{equation}
so that
\begin{equation}
        \int {dx\,d^2k_\bot\over 2(2\pi)^3}~|\phi(x,k_\bot)|^2 = 1. 
\label{momnor}
\end{equation}

\vskip 0.3cm
In principle, the momentum distribution amplitude
$\phi(x,k_\bot)$ can be obtained by solving the light-front
QCD bound state equation\cite{Zhang,Cheung}.
However, before such first-principles
solutions are available, we would have to be contented with
phenomenological amplitudes.  One example that has been often
used in the literature for heavy mesons is the so-called Bauer-Stech-Wirbel
(BSW) amplitude\cite{BSW}, which for a meson of mass $M$ is given by
\begin{equation}
        \phi(x,k_\bot)_{\rm BSW} = {\cal N} \sqrt{x(1-x)}
                ~{\rm exp}\left({-k^2_\bot\over2\omega^2}\right)
                ~{\rm exp}\left[-{M^2\over2\omega^2}(x-x_0)^2\right],
                \label{bswamp}
\end{equation}
where ${\cal N}$ is a normalization constant, $x$ is the
longitudinal momentum fraction carried by the light antiquark,
$x_0=({1\over2}-{m_1^2-m_2^2\over2M^2})$, 
and $\omega$ is a parameter of order $\Lambda_{\rm QCD}$.

   An other example is the Gaussian-type wave function,
\begin{equation}
        \phi(x,k_\bot)_{\rm Gauss}={\cal N} \sqrt{{dk_z\over dx}}
        ~{\rm exp}\left(-{\vec k^2\over 2\omega^2}\right),
        \label{gauss}
\end{equation}
where ${\cal N}=4(\pi/\omega^2)^{3/4}$, and $k_z$ of the internal momentum
$\vec k=(\vec{k}_\bot, k_z)$ is defined through
\begin{equation}
x = {e_1-k_z\over e_1 + e_2}, \qquad
1-x = {e_2+k_z \over e_1 + e_2},
\end{equation}
with $e_i = \sqrt{m_i^2 + \vec k^2}$. We then have 
\be
M_0=e_1 + e_2,~~~~k_z = \,{xM_0\over 2}-{m_2^2+k_\perp^2 \over 2 xM_0},
\label{kz}
\en
and
\begin{equation}
        {{dk_z\over dx}} = \,{e_1 e_2\over x(1-x)M_0}
\end{equation}
is the Jacobian of transformation from $(x, k_\bot)$ to
$\vec k$.
This wave function has been also used in many other studies
of hadronic transitions.
In particular, with appropriate parameters, it
describes satisfactorily the pion elastic form factor up
to $Q^2\sim 10~{\rm GeV}^2$ \cite{Chung}. A variant of the Gaussian-type
wave function is
\be
\phi(x,k_\perp)=\,{\cal N}\sqrt{dk_z\over dx}\exp\left(-{M_0^2\over 2\omega^2}
\right), \label{variant}
\en
with $M_0$ being given by (\ref{M0}). This amplitude is equivalent to
$\phi(x,k_\perp)_{\rm Gauss}$ when the constituent quark masses are equal but 
becomes different otherwise. Nevertheless, we will not pursue this wave 
function further because it does not have an appropriate heavy-quark-limit 
behavior (see Sec.~III). 

Obviously, the
Isgur-Wise function for heavy meson transitions depends
on the heavy meson wave function $\phi(x,k_\perp)$ chosen. It turns out that
in contrast to the Gaussian-type wave function, the BSW wave function 
fails to give a correct normalization for the
Isgur-Wise function at zero recoil in $P\to V$ transition.

\subsection{Decay Constants}
   The decay constant of a pseudoscalar meson $P(q_1\bar{q}_2)$ defined by
$\la 0|A^\mu|P\ra=\,if_{_P}p^\mu$
can be evaluated using the light-front wave function given by (2.1) and (2.5)
\be
\la 0|\bar{q}_2\gamma^+\gamma_5q_1|P\ra &=& \int \{d^3p_1\}\{d^3p_2\}
2(2\pi)^3\delta(P-p_1-p_2)\phi_P(x,k_\perp)R^{00}_{\lambda_1\lambda_2}
(x,k_\perp)   \non \\
&& \times\,\la 0|\bar{q}_2\gamma^+\gamma_5q_1|q_1\bar{q}_2\ra.
\en
Since $\widetilde{M}_0\sqrt{x(1-x)}=\sqrt{{\cal A}^2+k^2_\perp}$,
it is straightforward to show that
\be
f_P=\,4{\sqrt{3}\over\sqrt{2}}\int {dx\,d^2k_\perp\over 2(2\pi)^3}\,\phi_P(x,
k_\perp)\,{{\cal A}\over\sqrt{{\cal A}^2+k_\perp^2}}, \label{fp}
\en
where
\be
{\cal A}=m_1x+m_2(1-x).
\en
Note that the factor $\sqrt{3}$ in (\ref{fp}) arises from the color factor
implicit in the meson wave function. 

   Likewise, for the vector-meson decay constant defined by
\be
\la 0|V^\mu|V\ra=\,f_VM_V\ep^\mu,
\en
is found to be
\be
f_V &=& 4{\sqrt{3}\over \sqrt{2}}\int {dx\,d^2k_\perp\over 2(2\pi)^3}\,{\phi
_V(x,k_\perp)\over\sqrt{{\cal A}^2+k^2_\perp}}\,{1\over M_V}\Bigg\{ x(1-x)
M_V^2+m_1m_2+k^2_\perp   \non \\
&& +\,{{\cal B}\over 2m_V}\left[ {m_1^2+k_\perp^2\over 1-x}-{m_2^2+k_\perp^2
\over x}-(1-2x)M_V^2\right]\Bigg\},   \label{fv}
\en
where
\be
{\cal B}=xm_1-(1-x)m_2,~~~~W_V=M_0+m_1+m_2.
\en
When the decay constant is known, it can be used to constrain the
parameters of the light-front wave function.

\subsection{Form Factors for $P\to P$ Transition}
    With the light-front wave functions given above, we will first calculate
the form factors for $P\to P$ transitions given by
\be
\la P_2|V^\mu|P_1\ra =\,f_+(q^2)(P_1+P_2)^\mu+f_-(q^2)(P_1-P_2)^\mu,
\en
where $V^\mu=\bar{q}_2\gamma^\mu q_1$. For later purposes, it is also
convenient to parametrize this matrix element in different forms:
\be
\la P_2|V^\mu |P_1\ra &=& \sqrt{M_1M_2}\,\Big[\,h_+(q^2)(v_1+v_2)^\mu+h_-(q^2)
(v_1-v_2)^\mu\Big],   \non \\
&=& F_1(q^2)\left(P_1^\mu+P_2^\mu-{M_1^2-M_2^2\over q^2}\,q^\mu\right)+F_0
(q^2){M_1^2-M_2^2\over q^2}\,q^\mu,  \label{ppform}
\en
with $v_i\equiv P_i/M_i$, and
\be
F_1(q^2)=\,f_+(q^2),~~~~~F_0(q^2)=\,f_+(q^2)+{q^2\over M_1^2-M_2^2}f_-(q^2).
\label{Ff}
\en
In the heavy-quark limit $M_{1,2}\to\infty$, heavy-quark symmetry 
requires that \cite{Neu94}
\be
h_+(q^2)=\xi(v_1\cdot v_2),~~~~h_-(q^2)=0,   \label{ppHQS}
\en
where $\xi(v_1\cdot v_2)$ is the universal Isgur-Wise function normalized to
unity at the point of equal velocities: $\xi(1)=1$. 
The form factors $F_1$ and $F_0$ are related to
the transition amplitude with the exchange of a vector ($1^-$) and a 
scalar $(0^+)$ boson in the $t$-channel, respectively. 

   As explained in the Introduction, we shall work in the frame where
$q_\perp=0$ so that $q^2=q^+q^-$ will cover the whole time-like region 
$q^2\geq 0$. Define $r\equiv P_2^+/P_1^+$ (it is denoted by $R$ in 
\cite{Cheung2}), then
\be
  q^2=(1-r)\left(M_1^2-{M_2^2\over r}\right).
\en
Consequently, for a given $q^2$, there are two solutions for $r$:
\be
r_{\pm}=\,{M_2\over M_1}\left(v_1\cdot v_2\pm\sqrt{(v_1\cdot v_2)^2-1}\right),
\label{y12}
\en
where $v_1\cdot v_2$ is related to $q^2$ by
\be
v_1\cdot v_2=\,{M_1^2+M_2^2-q^2\over 2M_1M_2}.
\en
The $+ (-)$ signs in (\ref{y12}) correspond to the daughter meson recoiling in 
the positive (negative) $z$-direction relative to the parent meson (call them 
the ``+" and ``$-$" reference frame, respectively). At zero recoil 
($q^2=q_{\rm max}^2$) and maximum recoil ($q^2=0$), 
$r_{\pm}$ are given by
\be
&& r_+(q_{\rm max}^2)=r_-(q_{\rm max}^2)=\,{M_2\over M_1},   \non \\
&& r_+(0)=1,~~~~~r_-(0)=\left({M_2\over M_1}\right)^2. \label{y12val}
\en
The form factors $f_\pm(q^2)$ of 
course should be independent of the reference frame chosen for the
moving direction of the daughter meson.
For a given $q^2$, suppose we obtain
\be
\la P_2|V^+|P_1\ra\Big|_{r=r_+}=2P_1^+H(r_+),~~~~\la P_2|V^+|P_1\ra\Big
|_{r=r_-}=2P_1^+H(r_-).  \label{mel1}
\en
It follows from (\ref{ppform}) that
\be
f_+(q^2) &=& {(1-r_-)H(r_+)-(1-r_+)H(r_-)\over r_+-r_-}\,,  \non \\
f_-(q^2) &=& -{(1+r_-)H(r_+)-(1+r_+)H(r_-)\over r_+-r_-}\,. \label{fpm}
\en
It is easily seen that $f_\pm(q^2)$ are independent of the choice of ``+" or
``$-$" frame, as it should be.

   As noted earlier, in a frame with $q^+>0$, there are actually two distinct
contributions to the hadronic matrix element \cite{Jaus,Jaus96,Dubin,Saw}:
valence (partonic)
contribution calculated with relativistic light-front bound-state wave 
functions, and non-valence (non-partonic) contribution (or the so-called 
$Z$-graph) arising from quark-antiquark pair creation from the vacuum. In the 
following, we shall first provide some details for calculating the valence 
contribution, and then come back to the
non-valence subprocess in Sec.~II.D. For $P_1=(q_1\bar q)$ 
and $P_2=(q_2\bar q)$, the relevant quark momentum variables are
\def\bq{{\bar{q}}}
\be
&&p_1^+=(1-x)P_1^+,~~~p^+_\bq=xP_1^+,~~~\vec{p}_{_{1\perp}}=(1-x)\vec{P}_{
1\perp}+\vec{k}_\perp,~~~\vec{p}_{_{\bq\perp}}=x\vec{P}_{1\perp}-\vec{k}_
\perp,\non \\
&& p_2^+=(1-x')P_2^+,~~~p'^+_\bq=x'P_2^+,~~~\vec{p}_{_{2\perp}}=(1-x')\vec{P}_
{2\perp}+\vec{k}'_\perp,~~~\vec{p'}_{_{\bq\perp}}=x'\vec{P}_{2\perp}-\vec{k}'_
\perp,  \label{transmom}
\en
where $x~(x')$ is the momentum fraction carried by the spectator antiquark
$\bq$ in the initial (final) state. The spectator model requires that
\be
p'^+_{\bq}=p_\bq^+,~~~~~\vec{p'}_{_{\bq\perp}}=\vec{p}_{_{\bq\perp}}. 
\label{momeq}
\en
Taking a Lorentz frame where $\vec{P}_{1\perp}=\vec{P}_{2\perp}=0$ amounts 
to having $\vec{q}_\perp=0$ and $\vec{k}'_\perp=\vec{k}_\perp$. Then we 
readily obtain
\be
        \langle P_2|V^+|P_1\rangle =\sum_{\lambda_1,\lambda_2,\bar \lambda}
                \int && \{d^3p_{_\bq}\}
                \phi^*_2(x',k_\bot) \phi_1(x,k_\bot) \nonumber\\
                && \times R^{00\dagger}_{\lambda_2\bar \lambda}(x',k_\bot)
                 ~ \bar u(p_2,\lambda_2)\gamma^+ u(p_1,\lambda_1)
                R^{00}_{\lambda_1\bar \lambda}(x,k_\bot), \label{mel2}
\en
Substituting the covariant form given in Eq. (\ref{covariant})
into Eq. (\ref{mel2}) yields
\be
        \langle P_2|V^+|P_1\rangle =\,\sqrt{1\over r}
                \int &&{dx \,d^2k_\bot \over2(2\pi)^3}\,
                \phi^*_2(x',k_\bot) \phi_1(x,k_\bot)
                {-1 \over 2 \widetilde M_{02} \widetilde M_{01}
                \sqrt{(1-x')(1-x)} } \nonumber\\
                && \times {\rm Tr}\left[\gamma_5(\not{\! p}_2+m_2)\gamma^+
                (\not{\! p}_1+m_1) \gamma_5 (\not{\! p}_\bq-m_\bq)\right].
        \label{mel3}
\en
After some manipulation, the trace term in the above expression is reduced to
\be
        {\rm Tr}\left[\gamma_5(\not{\! p}_2+m_2)\gamma^+ (\not{\! p}_1
                +m_1) \gamma_5 (\not{\! p}_\bq-m_\bq)\right] 
=-{4\over x'}({\cal A}_1{\cal A}_2+k^2_\perp)P^+_1,   \label{trace}
\en
where
\be
{\cal A}_1=\,m_1x+m_\bq(1-x),~~~~~{\cal A}_2=\,m_2x'+m_\bq(1-x'), \label{a12}
\en
and use of (\ref{transmom}) has been made. Since
\be
\widetilde{M}_{01}\sqrt{x(1-x)}\,\widetilde{M}_{02}\sqrt{x'(1-x')}=
\sqrt{{\cal A}_1^2+k_\perp^2}\sqrt{{\cal A}_2^2+k_\perp^2},  \label{den}
\en
we find from (\ref{mel1}), (\ref{mel2}), (\ref{trace}) and (\ref{den}) that
\be
H(r)=\int^{r}_0dx\int {d^2k_\perp\over 2(2\pi)^3}\,\phi^*_2(x',k_\perp)
\phi_1(x,k_\perp)\,{{\cal A}_1{\cal A}_2+k^2_\perp\over\sqrt{{\cal A}^2_1
+k_\perp^2}\sqrt{{\cal A}_2^2+k_\perp^2}},   \label{Hi}
\en
with $x'=x/r$. The form factors $f_\pm(q^2)$ can then be obtained from 
(\ref{fpm}).

  As stated before, in the literature these form factors are customarily 
evaluated in the frame where $q^+=P_1^+-P_2^+=0$. This leads to 
$q^2=-q^2_\perp\leq 0$, implying a space-like momentum transfer. The 
advantage of the condition $q^+=0$ is that form factors only receive
valence contributions (see Sec.~II.D). However, there are two drawbacks in 
this approach: 
First, form factors in the physical time-like region cannot be obtained
without making additional $q^2$ extrapolation assumptions. Second, no 
information can be 
obtained for the form factor $f_-(q^2)$ since
$P_2^+=P_1^+$ [see (2.26)]. At the maximum recoil $q^2=0$, the form factor 
$f_+(0)$ is 
evaluated to be \cite{Jaus,Don95}
\be
f_+(0)=\int^1_0dx\int {d^2k_\perp\over 2(2\pi)^3}\,\phi^*_2(x,k_\perp)
\phi_1(x,k_\perp)\,{{\cal A}_1{\cal A}_2+k^2_\perp\over\sqrt{{\cal A}^2_1
+k_\perp^2}\sqrt{{\cal A}_2^2+k_\perp^2}},   \label{fp0}
\en
with ${\cal A}_1$ and ${\cal A}_2$ given by (\ref{a12}) except for that
$x'=x$ here. Therefore, the results of (\ref{fpm}) and (\ref{Hi}) at
$q^2=0$ are in agreement with (\ref{fp0}).

\subsection{Form Factors for $P\to V$ Transition}
   Form factors for $P\to V$ transition are defined as
\be
&& \la V(P_V,\ep)|J_\mu|P(P_1)\ra =\,{2\over M_P+M_V}i\epsilon_{\mu\nu\alpha
\beta}\ep^\nu P_V^\alpha P_1^\beta V(q^2)-\Big[ (M_P+M_V)\ep_\mu A_1(q^2)
\non \\
&&-{\ep\cdot P_1\over M_P+M_V}(P_1+P_V)_\mu A_2(q^2)-2M_V{\ep\cdot P_1\over
q^2}\,q_\mu\left(A_3(q^2)-A_0(q^2)\right)\Big],   \label{vpform}
\en
where $J_\mu\equiv V_\mu-A_\mu$, $A_3(0)=A_0(0)$, 
\be
A_3(q^2)=\,{M_P+M_V\over 2M_V}A_1(q^2)-{M_P-M_V\over 2M_V}A_2(q^2), 
\label{A312}
\en
and
\be
\ep^\mu(\pm 1)= \left( {2\vec{\ep}_\perp\cdot\vec{P}_{_{V\perp}}\over P_V^+},
\,0\,,\vec{\ep}_\perp\right),~~~~
\ep^\mu(0) &=& {1\over M_V}\left( {-M_V^2+P_{_{V\perp}}^2\over P_V^+},
\,P^+_V,\,\vec{P}_{_{V\perp}}\right)   \label{polvec}
\en
are, respectively, the transverse and longitudinal polarization vectors of 
the vector meson. The form factors $A_1$ and $A_2$ are related to $1^+$ 
intermediate states, $A_0$ to $0^+$ states, and $V$ to $1^-$ states. The
$P\to V$ matrix element can also be parametrized in different ways:
\be
\la V|J_\mu| P\ra &=& ig(q^2)\ep_{\mu\nu\alpha\beta}\ep^\nu P_V^\alpha
P_1^\beta+f(q^2)\ep_\mu+(\ep\cdot P_1)[a_+(q^2)(P_1+P_V)_\mu+a_-(q^2)(P_1-
P_V)_\mu]    \non \\
&=& \sqrt{M_PM_V}\Bigg\{i\tilde{g}(q^2)\epsilon_{\mu\nu\alpha\beta}\ep^\nu
v'^\alpha v^\beta+\tilde{f}(q^2)\ep_\mu  \non \\
&&~~~~~~~~~~~~~~+(\ep\cdot v)[\tilde{a}_+(q^2)(v+v')_\mu
+\tilde{a}_-(q^2)(v-v')_\mu]\Bigg\},  \label{vpform1}
\en
where $v=P_1/M_P$ and $v'=P_V/M_V$.
They are useful for later discussions. The form factors $a_\pm,~f$ and $g$ are
related to $V,~A_{0,1,2,3}$ via
\be
 g(q^2) &=& {2\over M_P+M_V}V(q^2),~~~f(q^2)=-(M_P+M_V)A_1(q^2),  \non \\
 a_+(q^2) &=& {1\over M_P+M_V}A_2(q^2),    \label{gV}  \\
 a_-(q^2) &=& {2M_V\over q^2}[\,A_3(q^2)-A_0(q^2)]   \non \\
 &=& {2M_V\over q^2}\left[ {M_P+M_V\over 2M_V}A_1(q^2)-{M_P-M_V
\over 2M_V}A_2(q^2)-A_0(q^2)\right].    \non    
\en
Also,
\be
\tilde{g}(q^2) &=& \sqrt{M_P+M_V}~g(q^2),~~~\tilde{f}(q^2)={f(q^2)\over
\sqrt{M_P+M_V}},   \\
\tilde{a}_+(q^2)+\tilde{a}_-(q^2)&=&{M_P^2\over{\sqrt{M_PM_V}}}\,(a_++a_-), 
~~~\tilde{a}_+(q^2)
-\tilde{a}_-(q^2)=\sqrt{M_PM_V}\,(a_+-a_-). \non 
\en
In the heavy-quark limit $M_P,M_V\to\infty$, heavy-quark symmetry demands
that \cite{Neu94}
\be
\tilde{a}_++\tilde{a}_-=0,~~~\tilde{a}_+-\tilde{a}_-=\tilde{g}=\xi(v\cdot
v'),~~~\tilde{f}=-(1+v\cdot v')\xi(v\cdot v'). \label{vpHQS}
\en

   The calculation of the $P\to V$ form factors is more subtle than the
$P\to P$ case. If we choose
a frame where $P_{1\perp}=P_{V\perp}=0$ as before, we will have $\ep\cdot
P_1=0$ for transverse polarization. As a result, form factors $a_\pm$ in 
(\ref{vpform1}) cannot be separately determined. Therefore, we will let 
$P_{1\perp}=P_{V\perp}{\ne}0$ at the outset, and set them to zero only
after the form factors are extracted. With the transverse polarization
$\ep^\mu(\pm 1)$, form factors $a_\pm(q^2)$ and $g(q^2)$ can be
individually determined. Then using the longitudinal polarization 
$\ep^\mu(0)$, we are able to fix the remaining form factor $f(q^2)$.

   We begin with $a_\pm(q^2)$. Since $\ep^+(\pm 1)=0$ [cf. (\ref{polvec})], it 
follows from (\ref{vpform1}) that
\be
-\la V(P_V)|A^+|P(P_1)\ra &=& (\ep\cdot P_1)[\,a_+(P_1^++P_V^+)+a_-(P_1^+
-P_V^+)] \non\\
&=& \left({{1\over r}-1}\right)(\vec{\ep}_\perp\cdot\vec{P}_\perp)[a_+(1+r)+a_-
(1-r)]P_1^+,    \label{vpme1}
\en
with $r\equiv P_V^+/P_1^+$ and $P_\perp\equiv P_{1\perp}=P_{V\perp}$. As will 
be shown below, the above matrix element at the quark level has the form
\be
\la V|A^+|P\ra=\,2\vec{\ep}_\perp\cdot\vec{P}_\perp(1-r)I(r)P^+_1.
\en
Substituting this into (\ref{vpme1}) and solving the equations for $r=r_+$ 
and $r=r_-$ yields
\be
a_+(q^2) &=& -{r_+(1-r_-)I(r_+)-r_-(1-r_+)I(r_-)\over r_+-r_-}\,,  \non\\
a_-(q^2) &=& {r_+(1+r_-)I(r_+)-r_-(1+r_+)I(r_-)\over r_+-r_-}\,, 
\label{apm}
\en
in analog to Eq.(\ref{fpm}) for $f_\pm(q^2)$.
In order to illustrate several subtle points in the derivation of $I(r)$,
we will go through the calculation in a bit more details. First of all, it is 
straightforward 
to show that for $P=(q_1\bar{q})$ and $V=(q_2\bar{q})$
\be
\la V|A^+|P\ra =\,\int{dx\,d^2k_\perp\over 2(2\pi)^3}\,{-2x'\over\sqrt{
{\cal A}_P^2+k^2_\perp}\sqrt{{\cal A}_V^2+k'^2_\perp}}\,\phi^*_V(x',k'_\perp)
\phi_P(x,k_\perp)(a+b),
\en
where ${\cal A}_P={\cal A}_1,~{\cal A}_V={\cal A}_2$ [see Eq.(\ref{a12})], and
\be
&& a=m_1(\hep\cdot p_2 p^+_\bq+\hep\cdot p_\bq p^+_2)+m_2(\hep\cdot p_1 p^+
_\bq-\hep\cdot p_\bq p^+_1)+m_\bq(\hep\cdot p_2 p^+_1+\hep\cdot p_1 p^+_2), 
\label{ab1}   \\
&& b={\hep\cdot(p_2-p_\bq)\over W_V}(m_1m_\bq p^+_2-m_2m_\bq p^+_1-m_1m_2
p^+_\bq+p_1\cdot p_\bq p^+_2-p_1\cdot p_2p^+_\bq+p_2\cdot p_\bq p^+_1), \non
\en
with $\hep^\mu=\ep^\mu(\pm 1)$ given by (\ref{polvec}) and $W_V=M_{0V}
+m_2+m_\bq$. By virtue of (\ref{transmom}) we find that
\be
a &=& (1-r)(1-2x'){\cal A}_P(\vec{\ep}_\perp\cdot\vec{P}_\perp)P_1^++\cdots,
\non \\
b &=& -2(1-r)\,{{\cal A}_P{\cal B}_V+k^2_\perp\over W_V}(\vec{\ep}_\perp
\cdot\vec{P}_\perp)P_1^++\cdots,   \label{ab2}
\en
with
\be
{\cal B}_V=-m_2x'+(1-x')m_\bq.
\en
The ellipses in (\ref{ab2}) denote contributions from terms proportional to 
$\vec{\ep}_\perp\cdot\vec{k}_\perp$ in (\ref{ab1}). Naively, these terms 
linear in $\vec{k}_\perp$ are not expected
to make contributions after integrating over $\vec{k}_\perp$. But this is not 
the case. Consider the term
\be
\widetilde{\phi}_V=\,{\phi_V(x',k'_\perp)\over\sqrt{{\cal A}_V^2
+k'^2_\perp}}
\en
and note that $k'_\perp$ is different from $k_\perp$ due to a non-vanishing
$P_\perp$:
\be
k'_\perp=\,k_\perp+(x'-x)P_\perp,  \label{kp}
\en
where we have used (\ref{transmom}) and (\ref{momeq}). Consequently,
\be
\widetilde{\phi}_V(k'^2_\perp) &=& \widetilde{\phi}_V(k^2_\perp)+(d\widetilde
\phi_V/dk^2_\perp)(k'^2_\perp-k^2_\perp)+\cdots    \non \\ 
&=& \widetilde{\phi}_V(k^2_\perp)\left[\,
1+2\Theta_V(x'-x)\kp+\cdots\right],   \label{thetav}  
\en
with 
\be
\Theta_V\equiv\,{1\over\widetilde \phi_V}\left({d\widetilde{\phi}_V\over 
dk^2_\perp}\right).                    
\en
Since
\be
\int d^2k_\perp(\ek)(\kp)=\,{1\over 2}\int d^2k_\perp k^2_\perp(\epp),
\en
it is evident that the linear term $(\ek)$ in (\ref{ab2}) will combine with
the linear term $(\kp)$ in (\ref{thetav}) to make a contribution to 
$\la V|A^+|P\ra$. We
wish to stress that this additional contribution from $\Theta_V$ was first
noticed and obtained by O'Donnell and Xu \cite{Don94a,Don95} and was neglected 
in the work of Jaus \cite{Jaus,Jaus96}.

   By the same token, in the expression of $b$, the $(\ek)$ term in 
$\hep\cdot(p_2-p_\bq)$ will also combine with the $(\kp)$ term in $(m_1m_\bq 
p_2^++\cdots)$ to
yield a contribution proportional to $\epp$ after integration over
$k_\perp$. The final result is
\be
I(r) &=& -\int^{r}_0 dx\int {d^2k_\perp\over 2(2\pi)^3}\,{x'\phi_V^*
(x',k_\perp)\phi_P(x,k_\perp)\over\sqrt{{\cal A}_P^2+k^2_\perp}\sqrt{{\cal 
A}_V^2+k^2_\perp}}\,\Bigg\{(1-2x'){\cal A}_P   \non \\
&+& [(1-2x'){\cal A}_P-{\cal A}_V]\Theta_V(x',k_\perp) k^2_\perp-2{({\cal 
A}_P{\cal B}_V+k^2_\perp)(1+\Theta_V(x',k_\perp)k^2_\perp)+{1\over 2}k^2_\perp
\over W_V}\Bigg\},   \label{Ii1}
\en
with $x'=x/r$.
In deriving (\ref{Ii1}) we have first integrated out $x'$ and $k'_\perp$. We 
can of course alter the order of integration by first integrating over $x$ and 
$k_\perp$ and obtain
\be
I(r) &=& -\int^1_0 dx'\int {d^2k_\perp\over 2(2\pi)^3}\,{x\phi_V^*
(x',k_\perp)\phi_P(x,k_\perp)\over\sqrt{{\cal A}_P^2+k^2_\perp}\sqrt{{\cal 
A}_V^2+k^2_\perp}}   \non \\
&\times& \Bigg\{ {\cal A}_V
 -[(1-2x'){\cal A}_P-{\cal A}_V]\Theta_P(x,k_\perp) k^2_\perp+2{({\cal A}_P
{\cal B}_V+k^2_\perp)\Theta_P(x,k_\perp)k^2_\perp+{1\over 2}k^2_\perp\over 
W_V}\Bigg\},   \label{Ii2}
\en
with $x=x'r$, where we have used the notation $k_\perp$ instead of 
$k'_\perp$ as it is a
dummy variable. The result (\ref{Ii2}) will be utilized in Sec.~3 to show that
light-front model calculations fulfill the heavy-quark-symmetry 
requirement (\ref{vpHQS}).

   At $q^2=0$, we have $r_+=1$ and $r_-=(M_V/M_P)^2$. It follows from 
(\ref{apm}) and (\ref{Ii1}) that
\be
A_2(0) &=& (M_P+M_V)a_+(0)=-(M_P+M_V)I(r=1)   \non \\
&=& \int^1_0 dx\int {d^2k_\perp\over 2(2\pi)^3}\,{x\phi_V^*
(x,k_\perp)\phi_P(x,k_\perp)\over\sqrt{{\cal A}_P^2+k^2_\perp}\sqrt{{\cal 
A}_V^2+k^2_\perp}}\,(M_P+M_V)\Bigg\{(1-2x){\cal A}_P    \\
&& +[(1-2x){\cal A}_P-{\cal A}_V]\Theta_V(x,k_\perp) k^2_\perp-2{({\cal A}_P
{\cal B}_V+k^2_\perp)(1+\Theta_V(x,k_\perp)k^2_\perp)+{1\over 2}k^2_\perp\over 
W_V}\Bigg\}.  \non
\en
This is in agreement with Eq.(30) of \cite{Don95}, but disagrees with the 
result obtained by Jaus \cite{Jaus}.

   Having fixed $a_\pm(q^2)$, we are ready to calculate $f(q^2)$ in 
(\ref{vpform1}). From (\ref{gV}) it is clear that once $f(q^2)$ 
is determined, so are the form factors $A_1(q^2)$ and $A_0(q^2)$. Since the 
``+" component of $\ep^\mu$ is needed to extract $f(q^2)$, we consider
the longitudinal polarization $\ep^\mu(0)$ of the vector meson $V$ and take a 
frame where $P_\perp=0$. Hence,
\be
-\la V|A^+|P\ra=\,f(q^2){r\over M_V}P_1^++{r\over 2M_V}\left(M_P^2-{M_V^2\over 
r^2}\right)[a_+(1+r)+a_-(1-r)]P^+_1.
\en
Let $\la V|A^+|P\ra\equiv J(q^2)P^+_1$, then
\be
f(q^2)=-{1\over 2}\left(M_P^2-{M_V^2\over r^2}\right)[a_+(1+r)+a_-(1-r)]-
{M_V\over r}J(q^2).   \label{fq2}
\en
After a straightforward manipulation, we obtain
\be
J(q^2)=\,-r\int_0^r dx\int {d^2k_\perp\over 2(2\pi)^3}\,{x'\phi_V^*
(x',k_\perp)\phi_P(x,k_\perp)\over\sqrt{{\cal A}_P^2+k^2_\perp}\sqrt{{\cal 
A}_V^2+k^2_\perp}}\,(c+d),   \label{jq2}
\en
with
\be
c &=& -{2\over M_{0V}}~ \left[ (1-x'){x'\over x}M_{0V}^2{\cal A}_P+{m_2m_\bq
\over x}{\cal A}_P+k_\perp^2(m_1+{m_2\over x}-m_\bq)\right],  
\non \\
d &=& {1\over M_{0V}}~ {1\over xW_V}\,({\cal A}_P{\cal B}_V+k^2_\perp)\left[
-(1-2x')M^2_{0V}+{m_2^2+k^2_\perp\over 1-x'}-{m_\bq^2+k^2_\perp\over x'}
\right], \label{cd}   
\en
and [cf. Eq.(\ref{M0})]
\be
M^2_{0V}=\,{m_2^2+k^2_\perp\over 1-x'}+{m_\bq^2+k^2_\perp\over x'}. 
\label{M0pv}
\en

   To check the above results, we note that for $r_+(0)=1$ 
[see (\ref{y12val})]
\be
\la V|A^+|P\ra \Big|_{r(0)=r_+(0)} &=& -[\,f(0)+(M^2_P-M^2_V)a_+(0)]P^+_1/M_V=
2A_0(0)P^+_1,
\en
so that 
\be
A_0(0) &=& {1\over 2}J(0)\Big|_{r=1}
\en
where use has been made of (\ref{gV}). Then it is not difficult to show from 
(\ref{jq2}-\ref{cd}) that 
\be
A_0(0) &=& \int {dx\,d^2k_\perp\over 2(2\pi)^3}\,{{\cal A}_P{\cal A}_V+(1-2x)
k^2_\perp+{2(m_1+m_2)xk^2_\perp\over W_V} \over \sqrt{{\cal A}_P^2+k^2_\perp}
\sqrt{{\cal A}_V^2+k^2_\perp} }\,
\en
which agrees with Eq.(27) of \cite{Don95} obtained in the
$q^+=0$ frame (implying $r=1$). It will be shown in Sec.~III.B that our 
result for $f(q^2)$ does respect the heavy-quark-symmetry requirement.

    Thus far we have imposed the condition $q_\perp=0$ to extract the form
factors $a_\pm(q^2)$ and $f(q^2)$. For the vector form factor $g(q^2)$ or
$V(q^2)$, it proves more convenient to first let $q_\perp\neq 0$ and then
set it to zero after the vector form factor is obtained. The ``+" 
component of the vector matrix element for transverse polarization reads
\be
\la V|V^+|P\ra &=& ig\epsilon_{+\nu\alpha\beta}\ep^\nu P_V^\alpha P_1^\beta
\non \\
&=& ig\epsilon_{+-xy}[\,-\ep^-q^x P^y+(P_1-P_V)^-\ep^x
P^y-P_1^-\ep^x q^y-(x\leftrightarrow y)],  \label{vme1}
\en
where $P_\perp\equiv P_{1\perp},~P_{V\perp}=P_\perp-q_\perp$. At the quark 
level, we have
\be
\la V|V^+|P\ra &=& \int ^r_0dx\int{d^2k_\perp\over 2(2\pi)^3}\,{2x'\phi_V^*
(x',k_\perp)\phi_P(x,k_\perp)\over\sqrt{{\cal A}_P^2+k^2_\perp}\sqrt{{\cal 
A}_V
^2+k^2_\perp}}\Bigg[ {\hep\cdot (p_2-p_\bq)\over W_V}i\epsilon_{+\alpha\beta
\gamma}p_2^\alpha p_1^\beta p_\bq^\gamma   \non \\
&+& i\epsilon_{+\alpha\beta\gamma}\hep^\alpha(m_1p_2^\beta p_\bq^\gamma
-m_2p_1^\beta p_\bq^\gamma+m_\bq p_2^\beta p_1^\gamma)\Bigg].   \label{vme2}
\en 
The transverse momentum variables are
\be
p_{1\perp}=\,(1-x)P_\perp+k_\perp,~~~p_{2\perp}=\,(1-x)P_\perp-q_\perp+k_
\perp,~~~p_{\bq\perp}=\,xP_\perp-k_\perp.
\en
It suffices to set $\hep^\mu=\ep^-(\pm)$ in (\ref{vme2}) to get contributions 
proportional
to $\ep_{+-xy}\ep^-(q^xp^y-q^yp^x)$, which is
related to the first term of (\ref{vme1}) at the hadron level.
It is easy to check that the transverse components $\hat\ep^x(\pm)$ and 
$\hat\ep^y(\pm)$ will not generate the same structure. Repeating the similar 
derivation as before, we obtain
\be
g(q^2) &=& {2V(q^2)\over M_P+M_V} =\int^r_0 dx\int{d^2k_\perp\over 2(2\pi)^3}
\,{2x'\phi_V^*(x',k_\perp)\phi_P(x,k_\perp)\over\sqrt{{\cal A}_P^2+k^2_\perp}
\sqrt{{\cal A}_V^2+k^2_\perp}} \non \\
&\times& \Bigg\{ {\cal A}_P+({\cal A}_P-{\cal A}_V)
\Theta_V(x',k_\perp) k^2_\perp   
+ {1\over W_V}\Big[ rk_\perp^2+(1-r)\left(2xM_{0P}k_z-{x'k^2_\perp\over 
1-x'}\right)   \label{gq2} \\   
&&~~~+(1-r)\Theta_V(x',k_\perp)k^2_\perp(2x^2M_{0P}^2-x'^2M_{0V}^2-m_{q'}^2-
k^2_\perp)\Big]\Bigg\},  \non 
\en
with $k_z$ being defined in (\ref{kz}). For $r(0)=r_+(0)=1$, 
(\ref{gq2}) leads to
\be
V(0) &=& \int^1_0 dx\int{d^2k_\perp\over 2(2\pi)^3}\, {(M_P+M_V)x\phi_V^*(
x',k_\perp)\phi_P(x,k_\perp)\over\sqrt{
{\cal A}_P^2+k^2_\perp}\sqrt{{\cal A}_V^2+k^2_\perp}} \non \\
&& \times\left({\cal A}_P+{
k_\perp^2\over W_V}+x(m_1-m_2)\Theta_V(x,k_\perp)k^2_\perp\right),
\en
which agrees with \cite{Don95}.

    Therefore, we have calculated the form factors $f(q^2),~g(q^2)$ and
$a_\pm(q^2)$ in the time-like $q^2$ region within the light-front framework.
Form factors $V(q^2)$ and $A_{0,1,2}(q^2)$ can then be determined via 
Eq.(\ref{gV}).\footnote{In the frame where $q^+=0$, only three of 
the $P\to V$ form factors, namely $f,~g$
and $a_+$ or $V,~A_1$ and $A_2$ are determined. However, $A_0$ can 
be fixed at $q^2=0$ using the relation
$A_0(0)=A_3(0)$ and (\ref{A312}).}

\subsection{Non-valence Contribution}
   Thus far we have concentrated on the valence-quark contribution to
the form factors. As stated in the Introduction, there also exist 
contributions which are generated from the quark-antiquark excitation or
higher Fock-states in the hadronic bound states. This additional $Z$-graph
contribution vanishes in the frame where the momentum transfer is 
purely transverse i.e., $q^+=0$, but survives otherwise.

   The general feature of the non-valence configuration can be recognized by
considering the quark triangle diagram (see Fig.~1). In terms of the ``+" 
component of momenta, the Feynman triangle diagram in the light-front framework
consists of two subprocesses: one corresponds to the 
valence-quark approximation for the meson wave functions, and the other
to the contribution of quark-pair creation from the vacuum. That is, through
the mechanism of quark-antiquark pair creation, the ``spectator" quark in 
the second subprocess is fragmented into a meson plus an outgoing quark. A 
detailed study of the quark triangle diagram for $P\to P$ transition 
gives (generalization to $P\to V$ transition is straightforward) \cite{Sima}
\be
\la P_2|\bar{q}_2\gamma^+q_1|P_1\ra =\,{\cal M}_a^++{\cal M}_b^+, 
\label{melMab}
\en
with
\be
{\cal M}_a^+ &=& g_1g_2\int^r_0{dx\over x(1-x)(1-x')}\int {d^2k_\perp\over 
2(2\pi)^3}\,{N_a^+\over (M_1^2-M_{01}^2)(M_2^2-M_{02}^2) },   \non \\
{\cal M}_b^+ &=& -g_1g_2\int^1_r{dx\over x(1-x)(1-x')}\int 
{d^2k_\perp\over 2(2\pi)^3}\,{N_b^+\over (M_1^2-M_{01}^2)(q^2-M_{12}^2){r\over 
1-r} },   \label{Mab}
\en
where $x'=x/r$, $g_1$ and $g_2$ are the quark-meson coupling constants
at different vertices, $M_1$ and $M_2$ are the masses of the initial and 
final meson respectively, $M_{01}^2~(M_{02}^2)$ is the same as
$M_{0P}^2~(M_{0V}^2)$ defined in (\ref{M0pv}), and
\be
M_{12}^2 &=& \left({m_1^2+k^2_\perp\over 1-x}-{m^2_2+k^2_\perp\over r-x}
\right)(1-r),   \non \\
N_a^+ &=& 4[\,p_1^+(m_2m_\bq+p_2\cdot p_\bq)+p_2^+(m_1m_\bq+p_1\cdot p_\bq)
+p^+_\bq(m_1m_2-p_1\cdot p_2)],   \\
N_b^+ &=& 4[\,p_1^+(-\eta M_1m_2+P_1\cdot p_2)+p_2^+(-\eta M_1m_1+
P_1\cdot p_1)+P_1^+(m_1m_2-p_1\cdot p_2)],   \non
\en
with $\eta=(m_1-m_\bq)/M_1$. Since ${\cal M}_a$ receives contributions from 
the kinematic region $0<x<r$ or $0<k^+<P_2^+$ (see Fig.~1), it corresponds 
to the valence-quark configuration.
As for ${\cal M}_b^+$, only the region $r<x<1$ or 
$P_2^+<k^+<P_1^+$ is relevant, and it corresponds to the non-valence
contribution. It is straightforward to check that, 
apart from a sign difference, $N_a^+$ is precisely the trace 
term given in (\ref{trace}). This implies that
the previous calculation for $P\to P$ form factors in the Hamiltonian
light-front approach is identical to the Feynman triangle graph under
the valence-quark approximation. Obviously, making the following substitutions
\be
{\sqrt{2}g_1\over x(1-x)}\,{1\over M_1^2-M_{01}^2}  ~&\longrightarrow&~
{\phi_1(x,k_\perp)\over\sqrt{{\cal A}_1^2+k^2_\perp}},   \non \\
{\sqrt{2}g_2\over x'(1-x')}\,{1\over M_2^2-M_{02}^2} ~ &\longrightarrow&~
{\phi_2(x',k_\perp)\over\sqrt{{\cal A}_2^2+k^2_\perp}}  
\en
in ${\cal M}_a^+$ will reproduce the result $\la P_2|\bar{q}_2\gamma^+q_1
|P_1\ra=2P_1^+H$ [see (\ref{mel1}) and (\ref{Hi})]. 

  Unlike the valence-quark contribution, only the wave function 
$\phi_1(x,k_\perp)$ of the initial meson enters into the expression of
${\cal M}_b^+$; $\phi_2(x',k_\perp)$ is not applicable for the non-valence
graph because the light-front momentum $k^+$ of the spectator quark is larger
than the momentum $P_2^+$ of the daughter meson (see Fig.~1).
This makes the task of calculating the effect of the $Z$-graph 
considerably more difficult. Nevertheless, some qualitative
features of ${\cal M}_b^+$ can still be comprehended. First of all, as noted 
earlier, the contribution from non-valence configurations vanishes in a 
frame where $q^+=0$ or $r=1$. 
However, this frame is suitable only for space-like $q^2$.
Second, it is easy to show that $N_b^+\to 0$ in the limit 
of heavy quark symmetry $m_Q\to\infty$, because it 
takes an infinite amount of energy to create a heavy quark-antiquark pair. 
This has the important implication that we do not have
to worry about the pair-creation subprocess when calculating the
Isgur-Wise function. Beyond the heavy-quark limit, it is commonly argued
that the non-valence contribution leads to a small correction in 
heavy-to-heavy transition but becomes more important for heavy-to-light decays
\cite{Dubin,Jaus96,Sima}. For example, a $B^*$-pole contribution is 
usually believed to be the dominant non-valence effect in $B\to\pi$ 
transition, especially when $q^2$ is near the zero-recoil point \cite{iw90}.
Some estimates based on the $B^*$-pole contribution with the help of
chiral perturbation theory indicate that for large values
of $q^2$, the $Z$-graph provides the dominant contribution to $B\to\pi$
form factors (for a recent estimate, see \cite{Cheung2}).

    In this paper we will demonstrate that even for heavy-to-heavy transition, 
the importance of the non-valence contribution depends on the
recoiling direction of the daughter meson.
As shown in (\ref{y12}), for a given $q^2$ there
are two possible reference frames characterized by $r_+(q^2)$ and $r_-(q^2)$,
corresponding to whether one chooses the velocity of the final
meson to be in the positive or negative $z$-direction relative to the initial 
meson. Of course, the form factors are independent of the choice of the
``+" or ``$-$" frame. This means that the combination 
${\cal M}_a^++{\cal M}_b^+$ in (\ref{melMab}) should be independent 
of the choice of $r(q^2)=r_+(q^2)$ or $r(q^2)=r_-(q^2)$.
From (\ref{Mab}) we see that the $r$
dependence of ${\cal M}_a^+$ or ${\cal M}_b^+$ appears both in the integrand
and in the integration limit. As a consequence, ${\cal M}_a^+$ and 
${\cal M}_b^+$ separately are in general ``$\pm$"-frame dependent. In
other words, {\it the valence-quark and non-valence contributions to
mesonic form factors are in general dependent on the recoiling direction
of the final meson, but their sum
is not}. For the form factors $f_\pm(q^2)$ in $P\to P$
decay and $a_\pm(q^2)$ in $P\to V$ transition, we have ``demanded" that
the valence contribution itself be
frame independent [see 
(\ref{mel1}), (\ref{fpm}) and (\ref{apm})]. For form factors $A_0,~A_1$
and $V$ in $P\to V$ decay, explicit calculations in Sec.~IV.B show that
the valence contributions for $r=r_+$ and $r=r_-$ are indeed different (see 
Fig.~6). Thus in principle we cannot make firm predictions for these form 
factors even for $B\to D^*$ transition, unless the non-valence contributions 
are also calculated. Nevertheless, corrections due 
to the non-valence configuration are expected to be marginal for
heavy-to-heavy form factors evaluated in the ``+" frame where $r=r_+$, but 
become more significant in the ``$-$" frame ($r=r_-$). The argument goes as 
follows: we know that the non-valence contribution vanishes if $q^+=0$. Now 
$q^+$ is never zero in the ``$-$" frame, whereas in the ``$+$" frame $q^+=0$ 
when $r_+=1$ [see (\ref{y12val})]. That means the valence-quark contribution
in the ``+" frame is {\it exact} at the $q^2=0$ point. As will 
be shown in Sec.~IV.B, the valence contributions at $q^2=0$
in the ``$-$" frame are generally smaller than those in the ``+" frame; the
difference should be accounted for by the non-valence configuration. These 
points will be elucidated in more detail in Sec.~IV.B.

\section{The Isgur-Wise Function}
   In Sec.~2 we have computed the $P\to P$ form factors $f_\pm(q^2)$ and 
$P\to V$ form factors $V(q^2),~A_{0,1,2}(q^2)$ for the entire physical $q^2$ 
region
using the light-front wave functions. It is very important to check if the
light-front model predictions are in accord with the requirements of 
heavy-quark symmetry, namely (\ref{ppHQS}) and (\ref{vpHQS}). In other words, 
as $m_Q\to\infty$, we would like to 
see if there exists a universal Isgur-Wise function which governs 
all heavy-to-heavy mesonic from factors in the light-front quark model.

  To our knowledge, the Isgur-Wise function has not been calculated 
directly for $q^2\geq 0$ within the framework of the light-front quark model,
though it has been considered in \cite{Don95,Simc,Mel}. The analysis of
\cite{Don95} is based on the observation \cite{Neu92} that the knowledge of
$P\to P$ or $P\to V$ form factors at $q^2=0$ (or at any point of $q^2$)
suffices to determine the Isgur-Wise function in the whole kinematic
region. However, this relies on the assumption that the model
calculations of form factors obey heavy-quark symmetry and that 
the universal form factor is only a function of $v\cdot v'$. The 
Isgur-Wise function is derived in \cite{Simc} from {\it space-like elastic}
form factors of heavy mesons,\footnote{This is based on the argument that, 
for the elastic form factor, 
$q^2=-(v\cdot v'-1)/(2M)$. Thus the space-like elastic form factor is related
to the Isgur-Wise function at time-like momentum transfers 
($v\cdot v'\geq 1)$.} while it is obtained in \cite{Mel} by
performing an analytic continuation from the region $q^2\leq 0$ to time-like 
momentum transfers. In contrast, we do not impose heavy-quark 
symmetry from the outset, so that we can check explicitly if the weak decay
form factors of heavy mesons can indeed be described by a single universal 
function when $m_Q\to \infty$. We will calculate this universal
function directly at the time-like momentum transfer to see if it is
independent of heavy quark masses and their ratio. It is important to note
that, since heavy quark-pair creation is forbidden in the $m_Q\to \infty$
limit, the $Z$-graph is no longer a problem in the reference frame where 
$q^+\geq 0$. Therefore, within the light-front quark model, we are able to 
compute the Isgur-Wise function $exactly$ for time-like $q^2$.

  To proceed,   
 we first investigate the heavy-quark-limit behavior of the wave function.
In the infinite quark mass limit $m_Q\to\infty$, the light-front wave 
function has the scaling behavior [17]:
\be
\phi_{Q_\bq}(x,k_\perp)\to~\sqrt{m_Q}\,\Phi(m_Qx,k_\perp),  \label{hqamp}
\en
where the factor $\sqrt{m_Q}$ or $\sqrt{M}$ ($M$ being the mass of the
heavy meson) comes from the particular normalization we have assumed
for the physical state in (\ref{wavenor}-\ref{momnor}). The reason why the 
light-front 
heavy-meson wave function should have such an asymptotic form is as follows.
Since $x$ is the longitudinal momentum fraction carried by the light 
antiquark, the meson wave function should be sharply peaked near $x\sim
\Lambda_{\rm QCD}/m_Q$. It is thus clear that only terms of the form
``$m_Qx$" survive in the wave function as $m_Q\to\infty$; that is,
$m_Qx$ is independent of $m_Q$ in the $m_Q\to\infty$ limit. For the BSW
wave function (\ref{bswamp}), we find that
\be
\Phi(X,k_\perp)_{\rm BSW}=\,\sqrt{32}\left({\pi\over \omega^2}\right)\exp
\left(-{k^2_\perp\over 2\omega^2}\right)\exp\left(-{X^2\over 2\omega^2}
\right)\sqrt{X},   \label{bswhq}
\en
where $X\equiv m_Qx$, and the normalization condition (\ref{momnor}) becomes
\be
\int^\infty_0dX\int {d^2k_\perp\over 2(2\pi)^3}\left|\Phi(X,k_\perp)
\right|^2=1.    \label{isnom}
\en
For the Gaussian-type wave function (\ref{gauss}), we obtain
\be
\Phi(X,k_\perp)_{\rm Gauss} =\, 4\left({\pi\over\omega^2}\right)^{3/4}\exp
\left(-{\vec{k}^2\over 2\omega^2}\right)\sqrt{{dk_z\over dX}}. \label{gausshq}
\en
From (\ref{kz}) it is clear that $k_z=[X-(m_\bq^2+k_\perp^2)/X]/2$ in the
heavy-quark limit. Therefore,
\be
\Phi(X,k_\perp)_{\rm Gauss} =\, 4\left({\pi\over\omega^2}\right)^{3/4}\exp
\left(-{k^2_\perp\over 2\omega^2}\right)\exp\left(-{({X\over 2}-{m_\bq^2+
k^2_\perp\over 2X})^2\over 2\omega^2}\right)\sqrt{{1\over 2}+
{m_\bq^2+k^2_\perp\over 2X^2}}.
\en
However, since $M_0\to m_Q+{\cal O}(m_Qx)$, it is clear that the wave function 
(\ref{variant}), which is a variant of the Gaussian type, does
not have the correct asymptotic form in the heavy-quark limit. Hence it is 
not suitable for describing heavy-quark transitions.

\subsection{$P\to P$ Transition in Heavy-Quark Limit}
    With the light-front wave function $\Phi(X,k_\perp)$ constructed in the 
$m_Q\to\infty$ limit, the $P\to P$ transition function $H(r)$ (\ref{Hi}) 
in the limit of heavy-quark symmetry (i.e., $m_1,m_2\to \infty$) becomes
\be
H(r)=\,\sqrt{M_2\over M_1}\int^\infty_0dX\int {d^2k_\perp\over 2(2\pi)^3}\,\Phi
(X',k_\perp)\Phi(X,k_\perp)\,{{\cal A}(X){\cal A}(X')+k^2_\perp\over\sqrt{
{\cal A}^2(X)+k^2_\perp}\sqrt{{\cal A}^2(X')+k^2_\perp}},
\en
where $X\equiv m_1x,~X'\equiv m_2x'$, and ${\cal A}(X)=
X+m_\bq$. Note that the quantities $X,~X',~m_\bq$ and 
$k_\perp$ appearing in the integrand are all
of order $\Lambda_{\rm QCD}$. Denote $z\equiv v_2^+/v_1^+=(M_1/M_2) r$, 
then [see (\ref{y12})],
\be
z_{\pm}=v_1\cdot v_2\pm\sqrt{(v_1\cdot v_2)^2-1}\,. \label{zpm}
\en
Obviously, $z_+z_-=1$ and $X'/X=1/z$.
Let $H(r_{\pm})=\sqrt{M_2/M_1}\,\widetilde{H}(z_{\pm})$,
so that (\ref{mel1}) can be rewritten as
\be
\la P_2|V^+|P_1\ra=2\sqrt{M_1M_2}\,\widetilde H(z)v_1^+.
\en
By a simple change of integration variable, one can readily show that 
\be
\widetilde H(z)=\,z\widetilde H(1/z).   \label{relhz}
\en

     To check the validity of the heavy-quark-symmetry relation (\ref{ppHQS}), 
we note that $h_\pm(q^2)$ are related to $\widetilde H(z)$ via
\be
h_{\pm}(q^2) &=& {1{\mp}z\over 1-z^2}\Bigg[\widetilde H(z){\pm}z \widetilde 
H(1/z)\Bigg],    \label{hpm}
\en
in analog to (\ref{fpm}) for $f_\pm(q^2)$. By virtue of (\ref{relhz}), the 
HQS relation $h_-(q^2)=0$ given in (\ref{ppHQS}) is indeed satisfied, and the 
Isgur-Wise function is given by
\be
\xi(v_1\cdot v_2)=\,{2\widetilde H(z)\over 1+z}.  \label{ppiw}
\en
Evidently, the Isgur-Wise function is independent of the heavy quark masses
$m_1,m_2$ and their ratio, but it depends on the light spectator quark mass.
The R.H.S. of (\ref{ppiw}) is invariant under the exchange $z \leftrightarrow
 1/z$, implying that the
Isgur-Wise function $\xi(v_1\cdot v_2)$ is independent of the choice of the
recoiling direction of the daughter meson, as it should be. At zero recoil
($z=1$), the expression for $\widetilde H(1)$ becomes identical to the 
normalization condition (\ref{isnom}). Hence $\widetilde H(1)=1$, and the 
Isgur-Wise function obeys the correct 
normalization condition $\xi(1)=1$. We would like to stress again that, 
unlike the previous works \cite{Don95,Mel} where $\xi(v\cdot v')$ is 
actually evaluated for $B\to D$ transition 
and for space-like values of $q^2$, here the Isgur-Wise function is obtained 
in the infinite quark mass limit and calculated directly for $q^2\geq 0$. 
Within the specific model we have taken, our result is exact. 
  
  In the limit of heavy-quark symmetry, form factors $F_1$ and $F_0$ are
related to the Isgur-Wise function via
\be
\xi(v_1\cdot v_2)=\,{2\sqrt{M_1M_2}\over M_1+M_2}\,F_1(q^2)=\,{2\sqrt{M_1
M_2}\over M_1+M_2}\,{F_0(q^2)\over \left[1-{q^2\over (M_1+M_2)^2}\right]}\,.
\label{ppiwform}
\en
Hence the $q^2$ dependence of $F_1$ is different
from that of $F_0$ by an additional pole factor.

\subsection{$P\to V$ Transition in Heavy-Quark Limit}
   There are four HQS relations given in (\ref{vpHQS}) for $P\to V$ form 
factors. We shall first focus on $\tilde{a}_\pm$ (or $a_\pm$). 
As $m_1,m_2\to\infty$, we can show that
\be
\Theta_V(x',k_\perp)\to \Theta(X',k_\perp),~~~~~\Theta_P(x,k_\perp)\to 
\Theta(X,k_\perp),
\en
with $\Theta_V$ being defined in (\ref{thetav}).
The $P\to V$ transition amplitude $I(r)$ [see (\ref{Ii1}) and (\ref{Ii2})]
reduce to
\be
I(r) &=& \widetilde{I}(z)=-{1\over\sqrt{m_1m_2}}\int {dXd^2k_\perp
\over 2(2\pi)^3}[X'{\cal A}(X)+X'(X-X')\Theta(X',k_\perp)k^2_\perp]f(X)f(X')
\non \\
&=& -{1\over\sqrt{m_1m_2}}\int {dX'd^2k_\perp
\over 2(2\pi)^3}[X{\cal A}(X')-X(X-X')\Theta(X,k_\perp)k^2_\perp]f(X)f(X'),
\en
where $f(X)=\Phi(X,k_\perp)/\sqrt{{\cal A}^2(X)+k^2_\perp},X'=X/z$, and all 
terms proportional 
to $1/W_V$ have been neglected in the heavy-quark limit. It is evident that 
$\widetilde I(z)$ satisfies the relation
\be
\widetilde I(z)=\,\widetilde I(1/z).
\en
Therefore, from (\ref{apm}),
\be
\tilde{a}_++\tilde{a}_-=\,{2M_P^2\over\sqrt{M_PM_V}}\,{r_+r_-\widetilde I
(z_+)-r_+r_-\widetilde I(z_-)\over r_+-r_-}=0,  \label{hqap}
\en
and
\be
\tilde{a}_+-\tilde{a}_-=\,2\sqrt{M_PM_V}\,{-r_+\widetilde I(z_+)+r_-\widetilde
I(z_-)\over r_+-r_-}=\,-2\sqrt{M_PM_V}\,\widetilde I(z). \label{hqam}
\en
By comparing this with (\ref{vpHQS}) yields the Isgur-Wise function 
\be
{\zeta}(v\cdot v') &=& 2\int^\infty_0 dX\int {d^2k_\perp\over 
2(2\pi)^3}\,{\Phi(X',k_\perp)\Phi(X,k_\perp)\over\sqrt{{\cal A}^2(X)+k^2_\perp}
\sqrt{{\cal A}^2(X')+k^2_\perp}}\,  \non \\
&\times& \left\{X'{\cal A}(X)+X'(X-X')\Theta(X',k_\perp)
k^2_\perp\right\}.    \label{vpiw} 
\en
with $X'/X=1/z$. It remains to show that $\zeta(v\cdot v')$ is indeed the 
same as the Isgur-Wise function $\xi(v\cdot v')$ found in $P\to P$ transition 
(\ref{ppiw}). We will address this 
issue later in Sec. IV. After showing the HQS relations (\ref{hqap}) and 
(\ref{hqam}) for form factors $\tilde{a}_\pm$, we turn to the vector form 
factor. One can easily show from (\ref{gq2}) that, indeed,
\be
\tilde{g}(q^2)=\,\sqrt{M_PM_V}\,g(q^2)~ \stackrel{\rm HQ~limit}{ 
\longrightarrow}~-2\sqrt{M_PM_V}\,\widetilde{I}(z)=\,\zeta(v\cdot v'),
\en
in accord with (\ref{vpHQS}).

  Using the results (\ref{hqap}) and (\ref{hqam}) for form factors 
$\tilde{a}_\pm$, we
are ready to prove the remaining HQS relation for $f(q^2)$. It follows from
(\ref{vpform}), (\ref{fq2}) and (\ref{jq2}-\ref{cd}) that
\be
\tilde{f}(q^2)={f(q^2)\over\sqrt{M_PM_V}}\,&=&-{1\over 2}\,{X^2-X'^2\over XX'}
\zeta+\int {dX\,d^2k_\perp\over 2(2\pi)^3}\,{\Phi(X',k_\perp)
\Phi(X,k_\perp)\over
\sqrt{{\cal A}_P^2+k_\perp^2}\sqrt{{\cal A}_V^2+k^2_\perp}}\left({x'M_V\over m_1}
c\right),
\en
where terms proportional to $1/W_V$ vanish in the limit of heavy-quark
symmetry. We find from (\ref{cd}) that
\be
{x'M_V\over m_1}c~\stackrel{{\rm HQ~limit}}{\longrightarrow}~
-2{X'\over X}\,({\cal A}_P{\cal A}_V+k^2_\perp),
\en
hence
\be
\tilde{f}(q^2)=-{1\over 2}\,{X^2-X'^2\over XX'}\zeta(v\cdot v')-{X'
\over X}\left(1+{X\over X'}\right)\xi(v\cdot v'),
\en
where use of (\ref{ppiw}) has been made. Then, using $X'/X=1/z$ and
(\ref{zpm}), we
are led to the desired HQS relation given in (\ref{vpHQS}):
\be
\tilde{f}(q^2)=-(1+v\cdot v')\xi(v\cdot v'),
\en
provided that $\zeta(v\cdot v')=\xi(v\cdot v')$.

  Is the function $\zeta(v\cdot v')$ given by
(\ref{vpiw}) identical to the Isgur-Wise function $\xi(v\cdot v')$ ? 
While $\xi(1)=1$ is always valid irrespective of the details of the 
light-front amplitude used,
the normalization of $\zeta(v\cdot v')$ at zero recoil is nontrivial. In fact,
we find that $\zeta(1)$ depends on the choice of the light-front
model wave function. We find numerically (see Sec.~IV.A) that the HQS 
requirement $\zeta(1)=1$ is
fulfilled by the Gaussian-type wave function (\ref{gauss}), 
but not so by the BSW-type wave function (\ref{bswamp}). 
In other words, {\it the normalization of the Isgur-Wise function at 
zero recoil in $P\to V$ transition puts a severe restriction
on the phenomenological light-front wave functions.} 
Since we are not able to solve the light-front QCD bound-state equation
to obtain the momentum distribution amplitude $\phi(x,k_\perp)$, we see
that heavy-quark symmetry is helpful in discriminating between different
phenomenological amplitudes. As will be shown in Sec.~IV.A,
${\zeta}(v\cdot v')$ is numerically equal to $\xi(v\cdot v')$ if the 
Gaussian-type wave function is used.
 
    The $P\to V$ form factors in the heavy-quark limit are all related to the 
Isgur-Wise function via
\be
\zeta(v\cdot v') &=& {2\sqrt{M_PM_V}\over M_P+M_V}\,V(q^2)=\,
{2\sqrt{M_PM_V}\over M_P+M_V}\,A_0(q^2)   \non \\
&=& {2\sqrt{M_PM_V}\over M_P+M_V}\,A_2(q^2)=\,
{2\sqrt{M_PM_V}\over M_P+M_V}\,{A_1(q^2)\over\left[1-{q^2\over (M_P+M_V)^2}
\right]}\,.    \label{HQSvp}
\en
That means $V,~A_0,~A_2$ all have the same $q^2$ dependence and they
differ from $A_1$ by an additional pole factor.

\section{Numerical Results and Discussions}
   To examine numerically the form factors derived in the last section, 
we need to specify the parameters appearing in the phenomenological 
light-front wave functions.
We shall use the decay constants to constrain the quark mass $m_q$ and
the scale parameter $\omega$. The decay constants of light pseudoscalar
and vector mesons are\footnote{The decay constant $f_\rho$ is obtained
from the measured decay rate of $\rho^0\to e^+ e^-$, while $f_{K^*}$ is 
determined from $\tau\to K^*\nu_\tau$.}
\be
f_\pi=\,132\,{\rm MeV},~~f_K=\,160\,{\rm MeV},~~f_\rho=\,216\,{\rm MeV},~~
f_{K^*}=\,210\,{\rm MeV}. \label{fpl}
\en
The decay constants of heavy mesons are unknown experimentally, so 
we have to rely on model calculations and lattice results. To be specific,
we take
\be
f_D=\,200\,{\rm MeV},~~f_B=\,185\,{\rm MeV},~~
f_{D^*}=\,250\,{\rm MeV},~~f_{B^*}=\,205\,{\rm MeV},    \label{fph}
\en
where the estimates for $f_{D^*}$ and $f_{B^*}$ are relatively more uncertain.
The parameters $m_q$ and $\omega$ in the Gaussian-type and BSW-type wave
functions fitted to the decay constants via (\ref{fp}) and (\ref{fv})
are listed in Table I. Note that the quark masses given in Table I are fixed 
to the commonly 
used values, and the other fitted values are by no means unique. Presumably, 
other hadronic properties, for example the light-meson
elastic form factor measured at a wide range of momentum transfer, would be 
helpful in fixing the light-front parameters.

\vskip 0.5cm
{\small Table I. Parameters $m_q$ (in units of GeV) and $\omega$ in the 
Gaussian-type and BSW-type wave functions fitted to the decay constants given 
by (\ref{fpl}) and (\ref{fph}).}
\begin{center}
\begin{tabular}{|c||c c c|c c c|c c c|c c c|} \hline
~wave function~ & $m_{u,d}$ & $\omega_\pi$ & $\omega_\rho$ & $m_s$ & 
$\omega_K$ 
& $\omega_{K^*}$ & $m_c$ & $\omega_D$ & $\omega_{D^*}$ & $m_b$ & $\omega_B$
& $\omega_{B^*}$ \\  \hline
Gaussian & ~0.25 & 0.33 & 0.30~ & ~0.40 & 0.38 & 0.31~ & ~1.6 & 0.46 & 0.47~ 
& ~4.8 & 0.55 & 0.55~ \\
BSW & ~0.25 & 0.30 & 0.29~ & ~0.40 & 0.34 & 0.30~ & ~1.6 & 0.46 & 0.46~ & ~4.8
& 0.58 & 0.57~ \\   \hline  
\end{tabular}
\end{center}
\vskip 0.5cm

\subsection{Results for the Isgur-Wise function}
   Before proceeding to numerically evaluate the $P\to P$ and $P\to V$ 
form factors, it is important to check the Isgur-Wise function to 
ensure that model calculations do respect heavy-quark symmetry in the 
infinite quark mass limit. With the Gaussian-type
(\ref{gausshq}) and BSW-type (\ref{bswhq}) wave functions given in
the limit of heavy-quark symmetry, the Isgur-Wise function $\xi(v\cdot 
v')$ for $P\to P$ transition calculated from (\ref{ppiw}) and (\ref{hpm}) is
shown in Fig.~2 using $\omega_D=\omega_B=0.55$. We see that the 
Isgur-Wise function obtained from Gaussian-type and BSW wave functions is
very similar. The slope of $\xi(v\cdot v')$ at the zero-recoil point is
\be
\rho^2\equiv-\xi'(1)=\,1.24\,.  \label{rho}
\en
Recent theoretical estimates and experimental analyses favor $\rho^2\lsim 1$. 
The slope parameter $\rho^2$ is subject to 
constraints from Bjorken and Voloshin sum rules (for a review, see 
\cite{Neu94}). A tight bound is derived to be $0.5<\rho^2<0.8$ \cite{Neu96}.
QCD sum-rule results range from 0.70 to 1.00 \cite{Neu96}. It thus appears
that our slope parameter (\ref{rho}) is too large. This may be attributed
to the fact that the Gaussian-type amplitude does not have enough amount
of high-momentum components at large $k_\perp$. It has been shown in
\cite{Simc} that the one-gluon-exchange interaction can generate 
high-momentum components in the meson wave function and reduce the value
of $\rho^2$ significantly.

   Although $\xi$ is independent of heavy quark masses, it is interesting to 
see if it can be
fitted to a simple pole behavior for a specific transition, e.g., $B\to D$~:
\be
\xi(q^2)=\,{\xi(0)\over(1-q^2/M_{\rm pole}^2)^\alpha}\,,
\en
where $v\cdot v'=(M_B^2+M_D^2-q^2)/(2M_BM_D)$. We find that $\xi(q^2)$ is 
fitted very well over the entire $q^2\geq 0$ region with a dipole behavior 
(one cannot tell the difference between fitted and calculated curves) with
\be
\alpha=2,~~~~~~M_{\rm pole}=\,6.65\,{\rm GeV}. \label{polefit}
\en
Indeed, this pole mass is close to the mass 6.34 GeV of the $1^+$ vector 
meson with $(b\bar c)$ content.

   The most interesting and striking results are shown in Fig.~3 for the 
function 
$\zeta(v\cdot v')$ [see (\ref{vpiw})] for
$P\to V$ transition obtained by taking the heavy-quark limit of the form 
factor $A_2(q^2)$ or $V(q^2)$. For the Gaussian-type wave function, 
we find that $\zeta(1)=1$ at zero recoil, and that numerically $\zeta(v\cdot 
v')$ is identical to 
$\xi(v\cdot v')$.\footnote{Since numerically $\zeta(v\cdot v')$ is equal to
$\xi(v\cdot v')$ up to six digits for the Gaussian-type amplitude, we believe
that this equivalence is {\it exact}, although both Maple and Methematica
fail to give an analytic result for (\ref{vpiw}).}
In contrast, the curve computed 
using the BSW amplitude deviates consistently from $\xi(v\cdot v')$; 
in particular, $\zeta(1)=0.87\,$ at zero-recoil. That means, for reasons 
not clear to us, the overlapping of the BSW wave functions for $P\to V$
transition at zero recoil is not complete in the heavy-quark limit.
This in turn implies that the
light-front amplitude $\Phi_{\rm BSW}$ is inconsistent with heavy-quark 
symmetry for $P\to V$ transition.

   We note that the presence of the $\Theta$ term in (\ref{vpiw})
is crucial for obtaining the numerical equivalence of $\zeta(v\cdot v')$ and
$\xi(v\cdot v')$. Hence the form factors 
$V(q^2),~A_1(q^2),~A_2(q^2)$ obtained previously in \cite{Jaus,Jaus96} are 
incomplete since the $\Theta$ terms are not taken into account there.

\subsection{$P\to P$ Form Factors}
  Since the BSW wave function fails to give a correct normalization at
zero recoil for the Isgur-Wise function in $P\to V$ transition, the ensuing
calculations are all carried out using the Gaussian-type wave function. The 
$q^2$ dependence of the form factors $F_1(q^2)=f_+(q^2)$ and $F_0(q^2)$ 
for $B\to D$ weak transition computed using
(\ref{Ff}), (\ref{fpm}) and (\ref{Hi}) are shown in Fig.~4 (we have neglected
the non-valence contributions). At $q^2=0$,
we obtain $F_1^{BD}(0)=F_0^{BD}(0)=0.70\,$. From Fig.~4 we see that 
$F_1^{BD}(q^2)$ can 
be fitted by a dipole approximation in the entire time-like $q^2$ region
with a pole mass $M_{\rm pole}=6.59$ GeV, in agreement with the pole mass 
6.65 GeV fitted to the Isgur-Wise function [cf. (\ref{polefit})],
while $F_0^{BD}(q^2)$ at low $q^2$
($0\leq q^2\lsim 6\,{\rm GeV}^2$) exhibits a monopole depenfence with 
$M_{\rm pole}=7.90$ GeV. This monopole behavior for $F_0^{BD}$ at low $q^2$
is consistent with (\ref{ppiwform}).

   The $q^2$ dependence of the form factor $f_+(q^2)$ for the transitions
$B\to\pi,~B\to K,~D\to \pi$ and $D\to K$ are shown in Fig.~5. The numerical
results for the form factors at $q^2=0$ are
\be
f_+^{B\pi}(0)=0.29\,,~~~~f_+^{BK}(0)=0.34\,,~~~~f_+^{D\pi}(0)=0.64\,,~~~~
f_+^{DK}(0)=0.75\,. \label{fpvalue}
\en
From Fig.~5 we see that, near the zero-recoil point,
the valence-quark prediction for $f_+^{B\pi}$ decreases as $q^2$ increases.
As explained in \cite{Cheung2}, the dipping of the valence-quark 
contribution toward the $q^2_{\rm max}$
point can be understood as follows.
Recall that the decay amplitude involves an overlapping integral of the
wave functions of the initial and final mesons. If both mesons were heavy, 
then it is obvious that, by heavy-quark symmetry, maximum overlapping 
must occur at the zero-recoil point. However, in the situation of $B\to\pi$
transition, the internal momentum distributions of the heavy $B$ meson 
and light pion peak at different values 
of $x$. Specifically, $\phi_B(x,k_\bot)$ has a narrow peak near $x=0$,
whereas $\phi_\pi(x,k_\bot)$ peaks with a much larger width at $x=1/2$.
Consequently, maximum overlapping of the wave functions actually
occurs somewhat away from the zero-recoil kinematics. For $D\to\pi$ 
transition, maximum overlapping occurs in the close vicinity of zero recoil 
(see Fig.~5). Since the non-valence contribution is expected to be important
for heavy-to-light form factors, especially for $B\to\pi$ transition,
a comparison with data at large $q^2$ cannot be made until such
contribution is included (form factors at $q^2=0$ are not affected by
the pair-creation configuration).\footnote{In \cite{Mel} form factors
at $q^2\leq 0$ are reformulated as a double dispersion integral 
representation, which allows one to perform an analytic continuation to
time-like momentum transfer. The Landau singularity there corresponds to our
valence-quark contribution, while the non-Landau singularity to the 
non-valence configuration. However, the contribution of the Landau
singularity in this approach vanishes at the ``quark zero recoil" point
(see Fig.~14 of \cite{Mel} for $D\to K$ transition), a phenomenon not seen
in our direct light-front calculations.}

   We see from (\ref{fpvalue}) that while the predicted $f_+^{DK}(0)$
is in nice agreement with experiment, $f_+^{DK}(0)_{\rm expt}=0.75
\pm 0.03$ \cite{PDG}; $f_+^{D\pi}(0)$ and the ratio $R=f_+^{D\pi}(0)/f_+^{DK}
(0)=0.87$
are too small compared to the measured values $R=1.29\pm 0.21\pm 0.11$ 
\cite{Ra} and $1.01\pm 0.20\pm 0.07$ \cite{Rb}. It has been pointed out in 
\cite{Chau} that the unexpected large decay rates of Cabibbo-suppressed 
decay $D^+\to\pi^+\pi^0$ and doubly-suppressed decay 
$D^0\to K^+\pi^-$ observed experimentally imply a sizeable SU(3)-breaking 
effect. This effect can be explained in the factorization approach only if 
$f_+^{D\pi}(0)>f_+^{DK}(0)$ or $R>1$. We find that explanation of the 
observed ratio $R$ remains an unsolved issue in the light-front 
quark model.

   Not shown in Fig.~5 is the heavy-to-light form factor $f_-(q^2)$, which
is expected to satisfy the heavy-quark-symmetry relation at $q^2$ near
zero recoil \cite{iw90}:
$(f_++f_-)^{B\pi}\sim 1/\sqrt{m_B}$ and $(f_++f_-)^{D\pi}\sim 1/\sqrt{m_D}$.
Our light-front calculation shows that in general $f_-(q^2)\sim -f_+(q^2)$ 
is a good approximation for $B(D)\to\pi$ transitions even when $q^2$ is
not close to $q^2_{\rm max}$, but it is only a rough approximation
for $B(D)\to K$ transitions.

\subsection{$P\to V$ Form Factors}
  The $q^2$ dependence of the form factors $V(q^2),~A_{0,1,2}(q^2)$ for
$B\to D^*$ transition is depicted in Fig.~6. We see that the valence-quark
contribution to $V,~A_0$ and $A_1$ depends on the choice of the ``+" or 
``$-$" reference frame, corresponding to $r(q^2)=r_+(q^2)$ or 
$r(q^2)=r_-(q^2)$. In general, the form factor in the ``+" frame
is larger than that in the ``$-$" frame, but they become identical at
zero recoil where $r_+(q^2_{\rm max})=r_-(q^2_{\rm max})=M_{D^*}/M_B$ 
[see (\ref{y12}-\ref{y12val})]. At maximum recoil $q^2=0$, we find
\be
V^{BD^*}(0)=\,0.78\,,~~~A^{BD^*}_0(0)=\,0.73\,,~~~
A^{BD^*}_1(0)=\,0.68\,,~~~A^{BD^*}_2(0)=\,0.61\,, \label{BDS0}
\en
in the ``+" frame where $r(0)=r_+(0)=1$, and
\be
V^{BD^*}(0)=\,0.62\,,~~~A^{BD^*}_0(0)=\,0.58\,,~~~
A^{BD^*}_1(0)=\,0.59\,,~~~A^{BD^*}_2(0)=\,0.61\,,~~~
\en
in the ``$-$" frame where $r(0)=r_-(0)=(M_{D^*}/M_B)^2$ [see (\ref{y12val})].
As discussed in Sec.~II.D, no firm
predictions for $V,~A_0,~A_1$ can be made unless the $Z$-graph contributions 
are included so that they are independent of the ``$\pm$"
frames. Although we do not have a reliable estimate for the
Z-graph contribution, we know that it is more 
important for the $r=r_-$ curve than the $r=r_+$ one. This is because 
form factors at $q^2=0$ do not receive the non-valence contribution in the 
``+" frame because $r_+(0)=1$. Therefore, (\ref{BDS0}) gives the complete
results for $B\to D^*$ form factors at $q^2=0$. Consequently, the 
difference between (\ref{BDS0}) and (4.8) must be equal to the non-valence 
contribution in the ``$-$" frame, namely,
\be
\widetilde V^{BD^*}(0)=\,0.16\,,~~~\widetilde A_0^{BD^*}(0)=\,0.15\,,~~~
\widetilde A_1^{BD^*}(0)=\,0.08\,,~~~\widetilde A_2^{BD^*}(0)=\,0\,.~~~
\en
This implies that {\it for heavy-to-heavy transition, form
factors calculated from the valence-quark configuration alone and evaluated 
in the ``+" frame 
should be reliable in a broad kinematic region and become
most trustworthy in the close vicinity of maximum recoil.} A generic
feature of the $Z$-graph effect is illustrated in Fig.~7 by considering
the form factor $A_0^{BD^*}$. {\it Assuming} that the full
$A_0^{BD^*}$ has a dipole behavior shown in Fig.~7 with a pole mass
$M_{\rm pole}=6.73$ GeV (dash-dotted curve), the difference between the ``full 
curve" and the valence contribution should give the non-valence contribution. 
It is clear that the $Z$-graph effect in the ``$-$" frame
(dashed curve) is sizeable in the entire kinemetic region, whereas it is 
important in the ``+" frame (solid curve) only when $q^2$ is close to the 
zero-recoil point.

  For a broad range of $q^2$, we find that $A_0^{BD^*},~A_2^{BD^*},~V^{BD^*}$
can be fitted to a dipole form and $A_1^{BD^*}$ to a monopole form, 
in accord with the HQS relations given in (\ref{HQSvp}).
   Experimentally, two form-factor ratios defined by
\be
R_1(q^2) &=& \left[1-{q^2\over (M_B+M_{D^*})^2}\right]\,{V^{BD^*}(q^2)
\over A_1^{BD^*}(q^2)},   \non \\
R_2(q^2) &=& \left[1-{q^2\over (M_B+M_{D^*})^2}\right]\,{A_2^{BD^*}
(q^2)\over A_1^{BD^*}(q^2)},   
\en
have been extracted by CLEO \cite{R12} from an analysis of angular 
distribution in $\bar{B}\to D^*\ell\bar{\nu}$ decays with the results:
\be
R_1(q^2_{\rm max})=\,1.18\pm 0.30\pm 0.12\,,~~~~R_2(q^2_{\rm 
max})=\,0.71\pm 0.22\pm 0.07\,.
\en
(\ref{HQSvp}) implies that, irrespective of the values of $q^2$,
 $R_1(q^2)=R_2(q^2)=1$ in the
heavy-quark limit. Our light-front calculations yield $V^{BD^*}
(q^2_{\rm max})=1.14\,,~A^{BD^*}_1(q^2_{\rm max})=0.83\,, $ and $~A^{BD^*}_2
(q^2_{\rm max})=0.96$, hence $R_1(
q^2_{\rm max})=1.11$ and $R_2(q^2_{\rm max})=0.92\,$, 
in agreement with experiment. The predictions of HQET are similar 
\cite{Neu94}: $R_1\simeq 1.3\pm 0.1$ and $R_2\simeq 0.8\pm 0.2\,$.

    As shown in Figs.~8-11, we have also computed the $q^2$ dependence of the 
form factors for $B\to K^*,~B\to\rho,~D\to K^*$ and $D\to\rho$ decays. The 
numerical results of the form factors 
at $q^2=0$ are (in the ``+" frame):
\be
&& B\to K^*:~~~~A_0^{BK^*}(0)=0.32\,,~~~A_1^{BK^*}(0)=0.26\,,~~~A_2^{BK^*}(0)=
0.23\,,~~~V^{BK^*}(0)=0.35\,,   \non \\
&& D\to K^*: ~~~~ A_0^{DK^*}(0)=0.71\,,~~~A_1^{DK^*}(0)=0.62\,,~~~A_2^{DK^*}
(0)=0.46\,,~~~V^{DK^*}(0)=0.87\,,    \non \\
&& B\to \rho:~~~~ A_0^{B\rho}(0)=0.28\,,~~~A_1^{B\rho}(0)=0.20\,,~~~A_2^{B
\rho}(0)=0.18\,,~~~V^{B\rho}(0)=0.30\,,    \\
&& D\to \rho:~~~~ A_0^{D\rho}(0)=0.63\,,~~~A_1^{D\rho}(0)=0.51\,,~~~A_2^{D
\rho}(0)=0.34\,,~~~V^{D\rho}(0)=0.78\,.   \non 
\en
Experimentally, only $D\to K^*$ form factors have been measured 
with the results \cite{PDG}
\be
V^{DK^*}(0)=1.1\pm 0.2\,,~~~~A_1^{DK^*}(0)=0.56\pm 0.04\,,~~~~A_2^{DK^*}(0)=
0.40\pm 0.08\,,  
\en
obtained by assuming a pole behavior for the $q^2$ dependence. Our predictions
for the $D\to K^*$ form factors are consistent with experiment. 

\vskip 0.5cm
\vfill\eject
\centerline{{\small Table II. Form factors for $B\to\rho$ and $D\to\rho$ 
transitions at $q^2=0$ in various models.}}
{\footnotesize \begin{center}
\begin{tabular}{|c|l|| c c c | c c c |} \hline
 & Reference & $A_1^{B\rho}(0)$ & $A_2^{B\rho}(0)$ & $V^{B\rho}(0)$ & 
$A_1^{D\rho}(0)$ & $A_2^{D\rho}(0)$ & $V^{D\rho}(0)$ \\  \hline
Lattice & BES \cite{BES} & -- & -- & -- & $0.65^{+15+24}_{-15-23}$ & 
$0.59^{+31+28}_{-31-25}$ & $1.07\pm 0.49\pm 0.35$    \\
   & LMMS \cite{LMMS} & -- & -- & -- & $0.45\pm 0.04$ & $0.02\pm 0.26$ & 
$0.78\pm 0.12$  \\
   & ELC \cite{ELC} & $0.22\pm 0.05$ & $0.49\pm 0.21\pm 0.05$ &
$0.37\pm 0.11$ & -- & -- & -- \\
  & APE \cite{APE} & $0.24\pm 0.12$ & $0.27\pm 0.80$ & $0.53\pm 0.31$ & --
   & -- & --  \\
  & UKQCD \cite{UKQCD} & $0.27^{+7+3}_{-4-3}$ & $0.28^{+9+4}_{-6-5}$  
& -- & -- & -- & -- \\
  & GSS \cite{GSS} & $0.16^{+4+22}_{-4-16}$ & $0.72^{+35+10}_{-35-7}$ &
$0.61^{+23+9}_{-23-6}$ & $~0.59^{+7+8}_{-7-6}$ & $~0.83^{+20+12}_{-20-8}$ 
& $~1.31^{+25+18}_{-25-13}~$  \\ 
\hline
Sum  & S \cite{S} & $0.96\pm 0.15$ & $1.21\pm 0.18$ & $1.27\pm 0.12$ &
-- & -- & -- \\
Rule  & Ball \cite{Ball} & $0.5\pm 0.1$ & $0.4\pm 0.2$ & $0.6\pm 0.2$ 
& $0.5\pm 0.2$ & $0.4\pm 0.1$ & $1.0\pm 0.2$ \\
   & ABS \cite{ABS} & $0.24\pm 0.04$ & -- & $0.28\pm 0.06$ & -- & -- & -- \\  
   & Narison \cite{Narison} & $0.38\pm 0.04$ & $0.45\pm 0.05$ & 
$0.45\pm 0.05$ & -- & -- & -- \\
   & YH \cite{YH} & $0.07\pm 0.01$ & $0.16\pm 0.01$ & $0.19\pm 0.01$ & 
$0.34\pm 0.08$ & $0.57\pm 0.08$ & $0.98\pm 0.11$ \\
   \hline
QM & ISGW \cite{ISGW} & 0.05 & 0.02 & 0.27 & -- & -- & -- \\
   & BSW \cite{BSW} & 0.28 & 0.28 & 0.33 & 0.78 & 0.92 & 1.23   \\  
  & Stech \cite{Stech} & 0.32 & 0.35 & 0.37 & -- & -- & -- \\ 
 & FGM \cite{FGM} & $0.26\pm 0.03$ & $0.31\pm 0.03$ & $0.29\pm 0.03$ & -- & 
 -- & -- \\
  & IV \cite{IV} & 0.50 & 0.51 & 0.70 & 0.55 & 0.45 & 1.08 \\
   \hline
LFQM &  {\bf this work} & {\bf 0.20} & {\bf 0.18} & {\bf 0.30} & {\bf 0.51} 
& {\bf 0.34} & {\bf 0.78}  \\
  & OXT \cite{Don95} & 0.21 & 0.18 & 0.32 & -- & -- & -- \\
  & Jaus \cite{Jaus96} & 0.26 & 0.24 & 0.35 & 0.58 & 0.42 & 0.93 \\
  & Melikhov \cite{Mel} & 0.17-0.26 & 0.16-0.24 & 0.22-0.34 & -- & -- & -- \\
   \hline
HQET & CDDGFN \cite{CDDGFNa} & 0.21 & 0.20 & 1.04 & 0.55 & 0.28 & 1.01 \\ 
+ChPT & CDDGFN \cite{CDDGFNb} & 0.28 & 0.19 & 0.50 & -- & -- & --   \\
\hline
\end{tabular}
\end{center}  }
\vskip 0.5cm

   Form factors for $B\to\rho$ and $D\to\rho$ transitions at $q^2=0$ 
predicted in various approaches (lattice simulations, QCD sum rule, 
quark model, light-front quark model, and heavy quark effective 
theory together with chiral perturbation theory) are summarized in 
Table II. We have to await further experimental studies in order to test 
various models. The $B\to K$ and
$B\to K^*$ transitions arise from flavor-changing neutral currents induced
by QCD corrections. It has been found recently that there are two
experimental data for $B\to J/\psi K^{(*)}$ which cannot be accounted for
simultaneously by all commonly used models \cite{Gour}.
Hence, it is important to have a reliable estimate of the $B\to K^{(*)}$ 
form factors at $q^2=m^2_{J/\psi}$ in order to
test the validity of the factorization approach. Our calculation gives
\be
F_1^{BK^*}(m^2_{J/\psi})=0.66\,,&&~~~~V_2^{BK^*}(m^2_{J/\psi})=0.42\,, \non\\
A_0^{BK^*}(m^2_{J/\psi})=0.63\,,&&~~~~A_1^{BK^*}(m^2_{J/\psi})=0.37\,,~~~~
A_2^{BK^*}(m^2_{J/\psi})=0.43\,,
\en
from valence-quark configuration.

   As for the $q^2$ dependence of heavy-to-light from factors, we see
from Figs.~8-11 that, except for $V^{B\rho}$,
they all increase with $q^2$, though $A_1$ is flatter than $A_0,~A_2$, and 
$V$. As we have argued before, the valence-quark
contribution evaluated in the ``+" frame should be reliable when $q^2$
is close to maximum recoil. For small $q^2$, we have a dipole behavior for
$A_0,~A_2,~V$ (except for $V^{B\rho}$ and $V^{BK^*}$)
and a monopole behavior for $A_1$; that is, $A_0,~A_2$, and$~V$ 
increase with $q^2$ faster than $A_1$. The form factor $V$ for $B\to\rho$ and
$B\to K^*$ in the ``+" frame do not have a dipole behavior at small $q^2$ 
mainly because of the large destructure contributions from the $\Theta_V 
k^2_\perp/W_V$ terms in (\ref{gq2}). As a result, the form factor $V$ 
in $B\to\rho$ and $B\to K^*$ decays evaluated in the ``+" frame is smaller
than that in the ``$-$" frame. The $q^2$ dependence of the $P\to V$
form factors have also been studied in the QCD-sum-rule approach with
some contradicting results. For example, while $A_1^{B\rho}$ is found to 
decrease from $q^2=0$ to $q^2=15\,{\rm GeV}^2$ in \cite{BBD} (see also
\cite{Ball,Narison,IV}), such a phenomenon is not seen in \cite{ABS,YH}
(see also Sec.~5.3 of \cite{CDDGFNb}). The sum-rule results of \cite{YH}
show that the form factors $A_0,~A_2,~V$ all
have a dipole form while $A_1$ has a monopole form, in accord with
our observation. The same conclusion is also reached in \cite{Don95}
based on the scaling behavior of heavy-to-light from factors in the 
$m_Q\to\infty$ limit. A recent lattice study of the axial form factors
$A_0^{B\rho},~A_1^{B\rho}$ and $A_3^{B\rho}$ \cite{UKQCD} is consistent
with the $q^2$ behavior we have in the light-front quark model.

\section{Summary}
  The heavy-to-heavy and heavy-to-light form factors in $P\to P$ and $P\to V$
transitions are studied in the present paper. In the light-front 
relativistic quark model, the decay form factors are evaluated in a frame
where $q^+\geq0$ and $q_\perp=0$, so that it covers the entire physical range 
of momentum transfer
and no extrapolation assumption from $q^2=0$ or from $q^2=q^2_{\rm max}$ 
is required. In previous works using $q^+=0$, one can only calculate form 
factors at $q^2$ =0; moreover, the form factors $f_-(q^2)$ in $P\to P$ decay 
and $a_-(q^2)$ in $P\to V$ decay cannot be studied. For the first time, we 
have calculated the $P\to V$ form factors directly at time-like momentum 
transfers.  The main results of this paper are :

   1). We have investigated the behavior of heavy-to-heavy form factors
in the heavy-quark limit and found that the requirements of 
heavy-quark symmetry (\ref{ppHQS}) for $P\to P$ transition and (\ref{vpHQS})
for $P\to V$ transition are indeed fulfilled by the light-front quark model
provided that the universal function $\zeta(v\cdot v')$ obtained from
$P\to V$ decay is identical to the Isgur-Wise function $\xi(v\cdot v')$
in $P\to P$ decay.

  2). Contrary to the Isgur-Wise function in $P\to P$ decay, the 
normalization of $\zeta(v\cdot v')$ at zero recoil depends on the light-front 
wave function used. We found that the BSW amplitude correctly gives
$\xi(1)=1$, but $\zeta(1)=0.87$. Therefore, this type of wave functions
cannot describe $P\to V$ decays in a manner consistent with heavy-quark
symmetry. 

   3). Using the Gaussian-type amplitude, the Isgur-Wise function $\zeta
(v\cdot v')$ has a correct normalization at zero recoil and is identical
to $\xi(v\cdot v')$ numerically up to six digits. It can be fitted very well
with a dipole dependence with $M_{\rm pole}=6.65$ GeV for $B\to D$ transition.
 However, the predicted slope parameter $\rho^2=1.24$ is probably too large.
This may be ascribed to the fact that the Gaussian-type wave function does 
not have enough high-momentum components at large $k_\perp$.

    4). The valence-quark and non-valence contributions to form factors 
are in general dependent on the recoiling direction of the daughter 
meson relative to the parent meson, but their sum should not. Although we do 
not have a reliable estimate of the pair-creation effect, we have 
argued that, for heavy-to-heavy transition, form factors calculated from the 
valence-quark configuration evaluated in the
``+" frame should be reliable in a broad kinematic region, and they become
most trustworthy in the vicinity of maximum recoil.

  5). The form factors $F_1,~A_0,~A_2,~V$ (except for $V^{B\rho}$ and 
$V^{BK^*}$) all exhibit a dipole behavior, which
$F_0$ and $~A_1$ show a monopole behavior in the close vicinity of maximum
recoil for heavy-to-light transition, and in a broader kinematic region
for heavy-to-heavy decays. Therefore, $F_1,~A_0,A_2,~V$ increase
with $q^2$ faster than $F_0$ and $A_1$.

\vskip 3.0cm
\centerline {\bf ACKNOWLEDGMENTS}

   This work was supported in part by the National Science Council of ROC
under Contract Nos. NSC85-2112-M-001-010 and NSC85-2112-M-001-023.

%
%
\newcommand{\bi}{\bibitem}
%

%
\newpage
\parindent=0 cm
\centerline{\bf FIGURE CAPTIONS}
\vskip 0.5 true cm

{\bf Fig. 1 } The Feynman triangle diagram and the corresponding light-front
subdiagrams. Diagram (a) corresponds to the valence-quark configuration and
diagram (b) to the non-valence configuration.
\vskip 0.25 true cm

{\bf Fig. 2 } The Isgur-Wise function $\xi(v\cdot v')$ for $P\to P$
transition calculated using Gaussian-type (solid line) and BSW-type (dashed
line) light-front 
wave functions. For comparison, a curve for $1/v\cdot v'$ is also shown.
\vskip 0.25 true cm

{\bf Fig. 3 } The Isgur-Wise function $\zeta(v\cdot v')$ for $P\to V$
transition calculated using Gaussian-type (solid line) and BSW-type (dashed
line) light-front wave functions. 
\vskip 0.25 true cm

{\bf Fig. 4 } Form factors $F_1(q^2)=f_+(q^2)$ and $F_0(q^2)$ for $B\to D$
transition arising from the valence-quark configuration.
Dashed curves are fits to $F_1$ in a dipole form with $M_{\rm pole}=6.59$ 
GeV and to $F_0$ in a monopole form with $M_{\rm pole}=7.90$ GeV.
\vskip 0.25 true cm

{\bf Fig. 5 } The form factor $f_+(q^2)$ for $B\to\pi,~B\to K,~D\to\pi$ and 
$D\to K$ transitions arising from the valence-quark configuration.
\vskip 0.25 true cm

{\bf Fig. 6 } Form factors $V(q^2),~A_0(q^2),~A_1(q^2)$ and $A_2(q^2)$ for
$B\to D^*$ transition. Solid lines are the valence contribution evaluated
in the ``+" frame where $r(q^2)=r_+(q^2)$, and dashed lines 
in the ``$-$" frame where $r(q^2)=r_-(q^2)$. The contribution to the form
factor $A_2$ is independent of the choice of the ``+" or ``$-$" frame.
\vskip 0.25 true cm

{\bf Fig. 7 } An illustration of the non-valence contribution to the 
form factor $A_0^{BD^*}$, whose general feature applies to $A_1^{BD^*}$  
and $V_0^{BD^*}$ as well. Valence-quark contributions to $A_0^{BD^*}$ 
evaluated in the ``+" frame (solid line) and in the ``$-$" frame 
(dashed line) are the same as in Fig.6. The corresponding non-valence 
contribution are extracted in respective frames by assuming that the 
full $A_0^{BD^*}$ has a dipole behavior with 
$M_{\rm pole}$ = 6.73 GeV (dash-dotted line).
\vskip 0.25 true cm

{\bf Fig. 8 } Same as Fig.~6 except for $B\to K^*$ transition.
\vskip 0.25 true cm

{\bf Fig. 9 } Same as Fig.~6 except for $B\to \rho$ transition.
\vskip 0.25 true cm

{\bf Fig. 10 } Same as Fig.~6 except for $D\to K^*$ transition.
\vskip 0.25 true cm

{\bf Fig. 11 } Same as Fig.~6 except for $D\to \rho$ transition.
\vskip 0.25 true cm


\begin{figure}[h]
\hskip 4cm
\hbox{\epsfxsize=8cm
      \epsfysize=20cm
      \epsffile{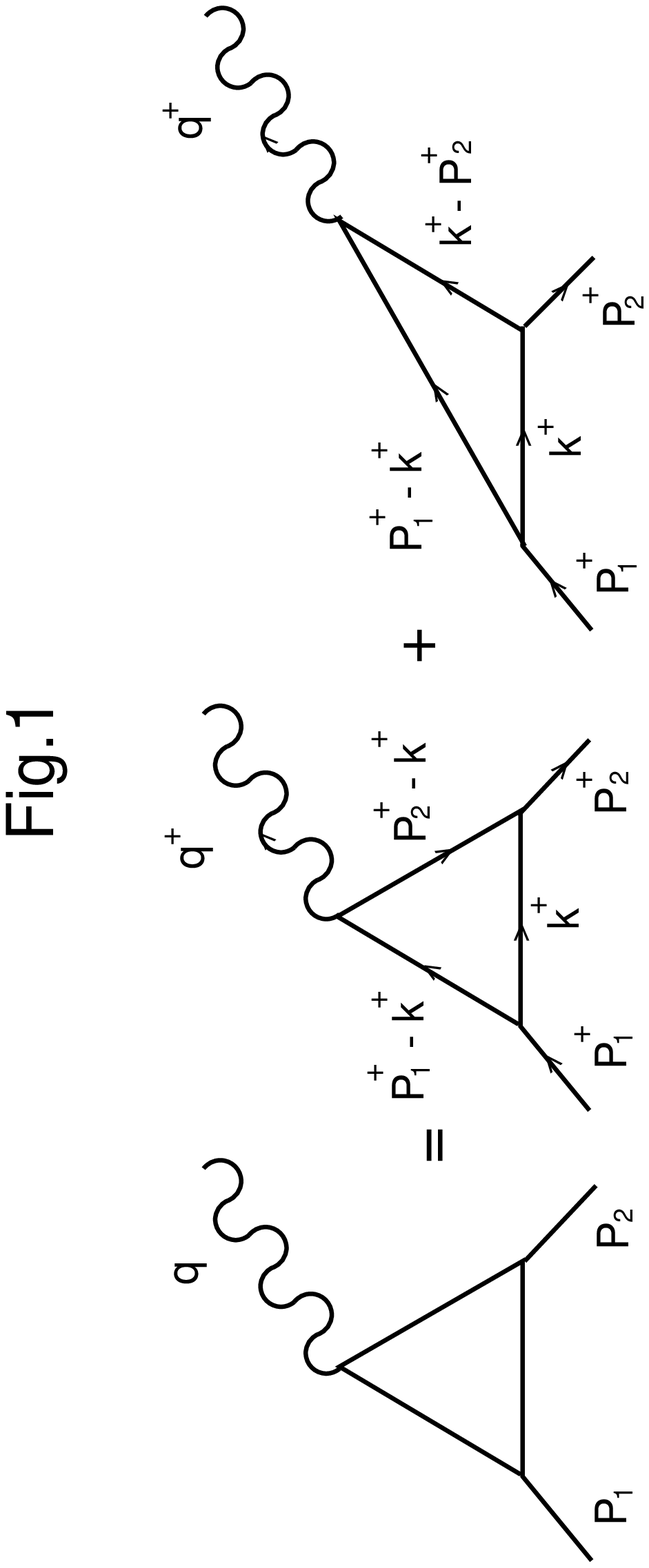}}
\end{figure}
\newpage

\begin{figure}[h]
\hskip 1cm
\hbox{\epsfxsize=16cm
      \epsfysize=20cm
      \epsffile{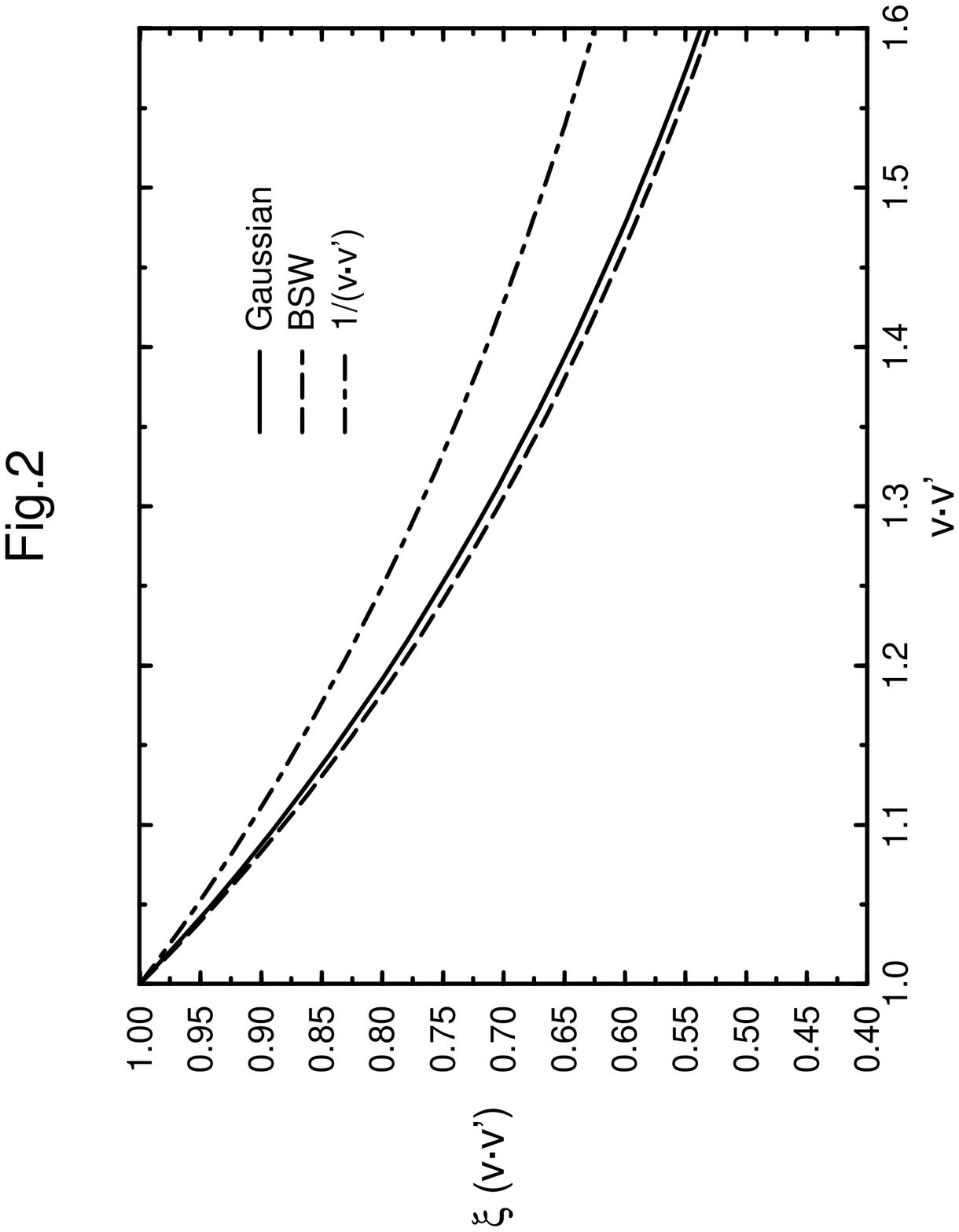}}
\end{figure}
\newpage

\begin{figure}[h]
\hskip 1cm
\hbox{\epsfxsize=16cm
      \epsfysize=20cm
      \epsffile{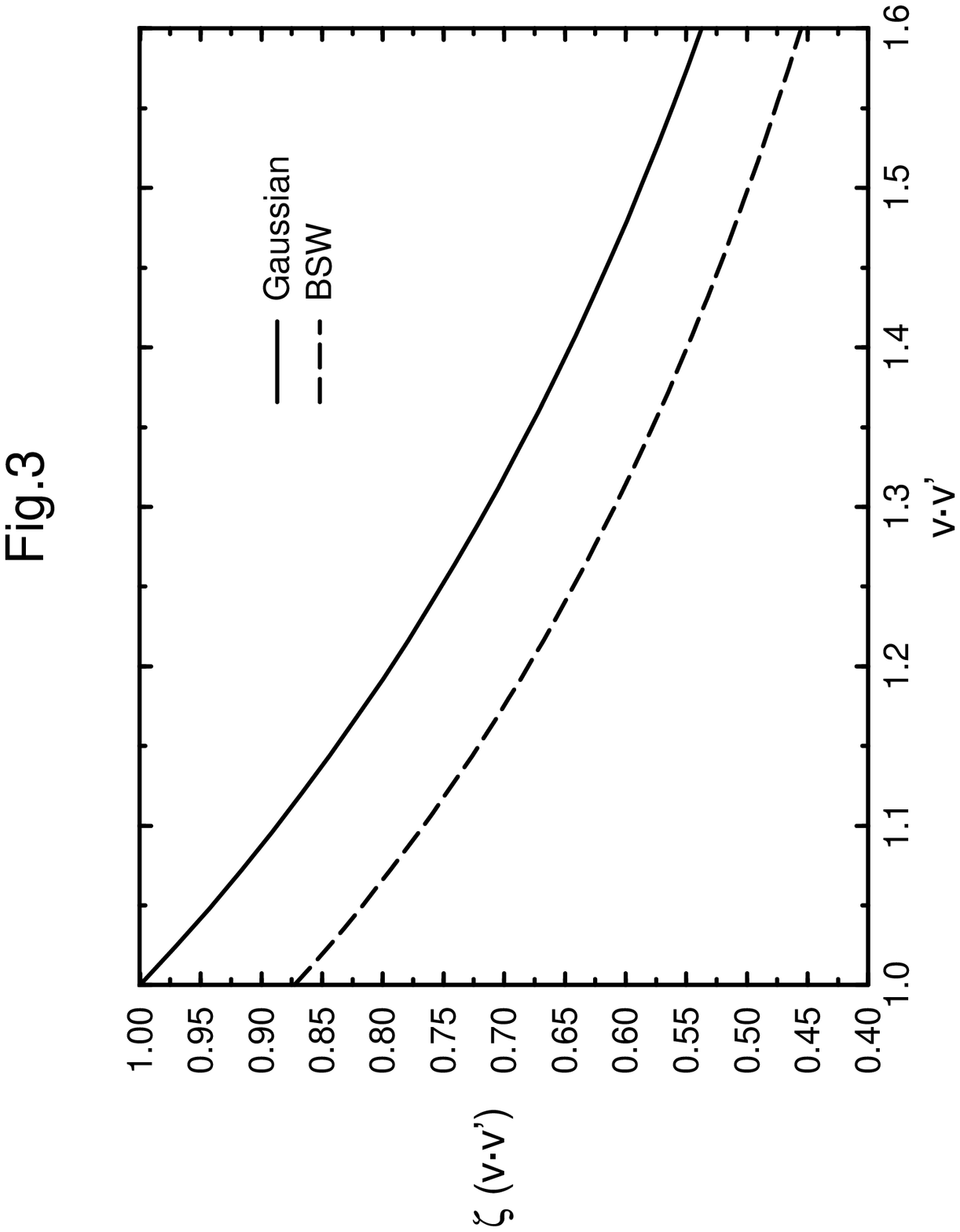}}
\end{figure}
\newpage

\begin{figure}[h]
\hskip 1cm
\hbox{\epsfxsize=16cm
      \epsfysize=20cm
      \epsffile{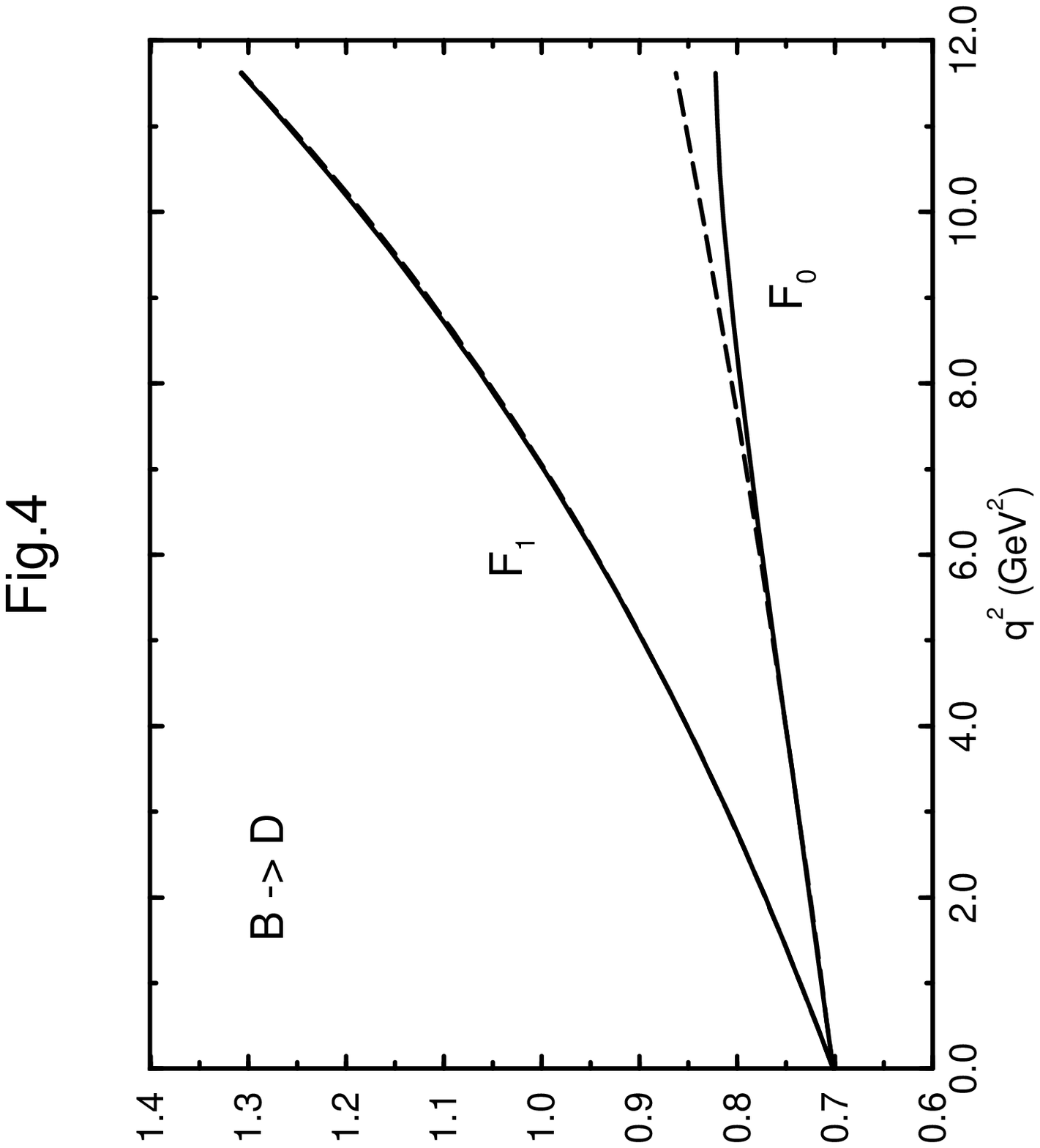}}
\end{figure}
\newpage

\begin{figure}[h]
\hskip 1cm
\hbox{\epsfxsize=16cm
      \epsfysize=20cm
      \epsffile{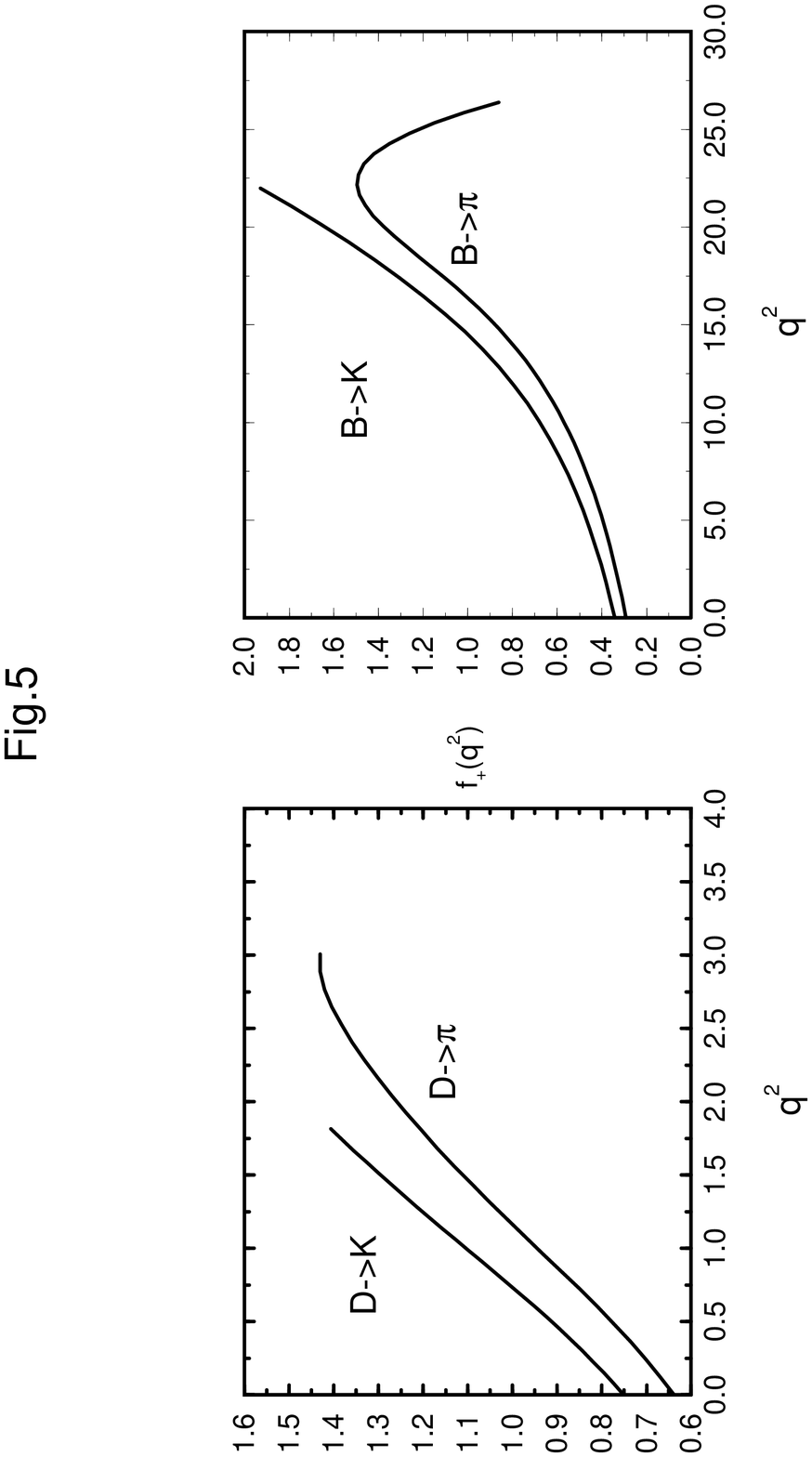}}
\end{figure}
\newpage

\begin{figure}[h]
\hskip 1cm
\hbox{\epsfxsize=16cm
      \epsfysize=20cm
      \epsffile{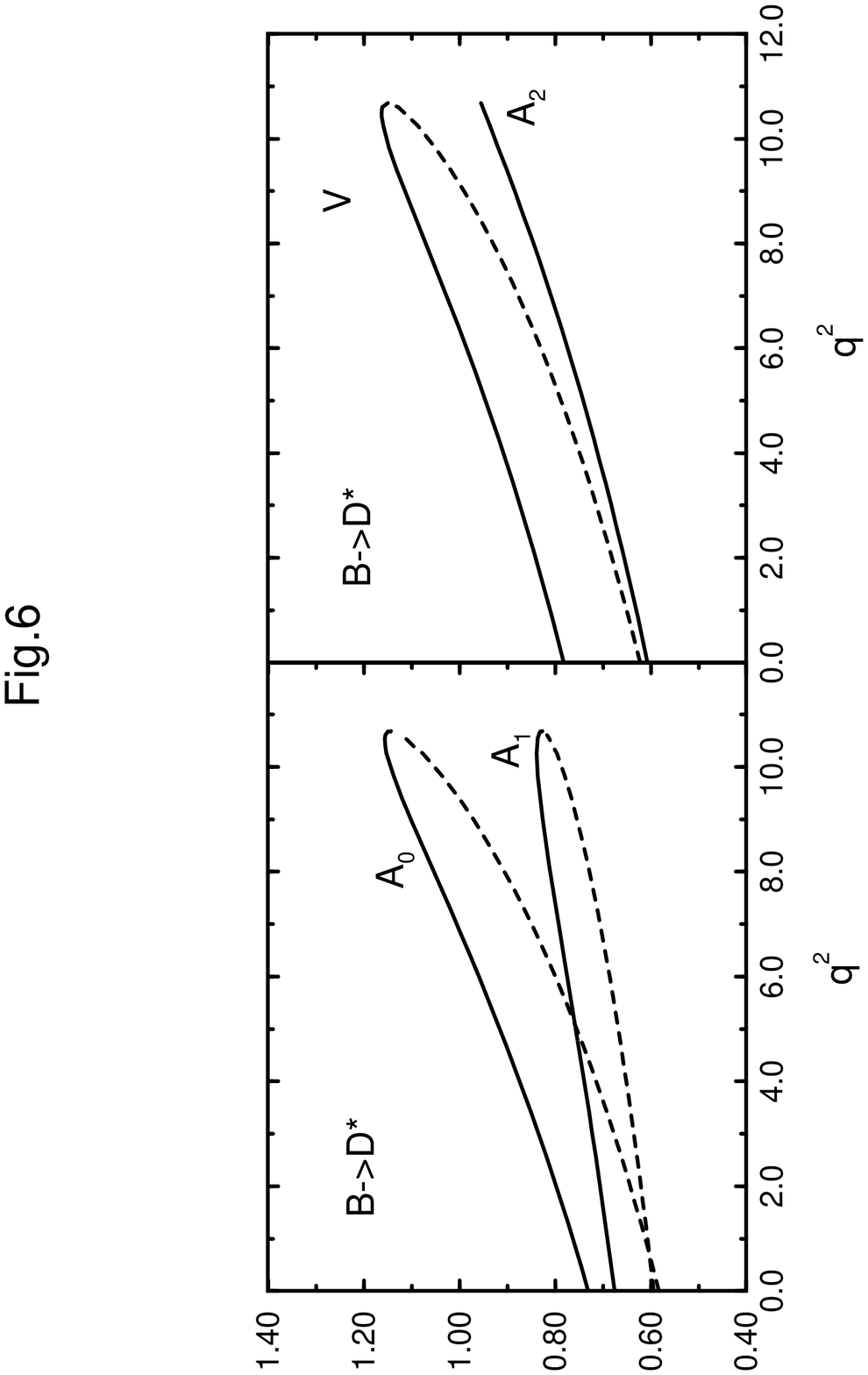}}
\end{figure}
\newpage

\begin{figure}[h]
\hskip 1cm
\hbox{\epsfxsize=16cm
      \epsfysize=20cm
      \epsffile{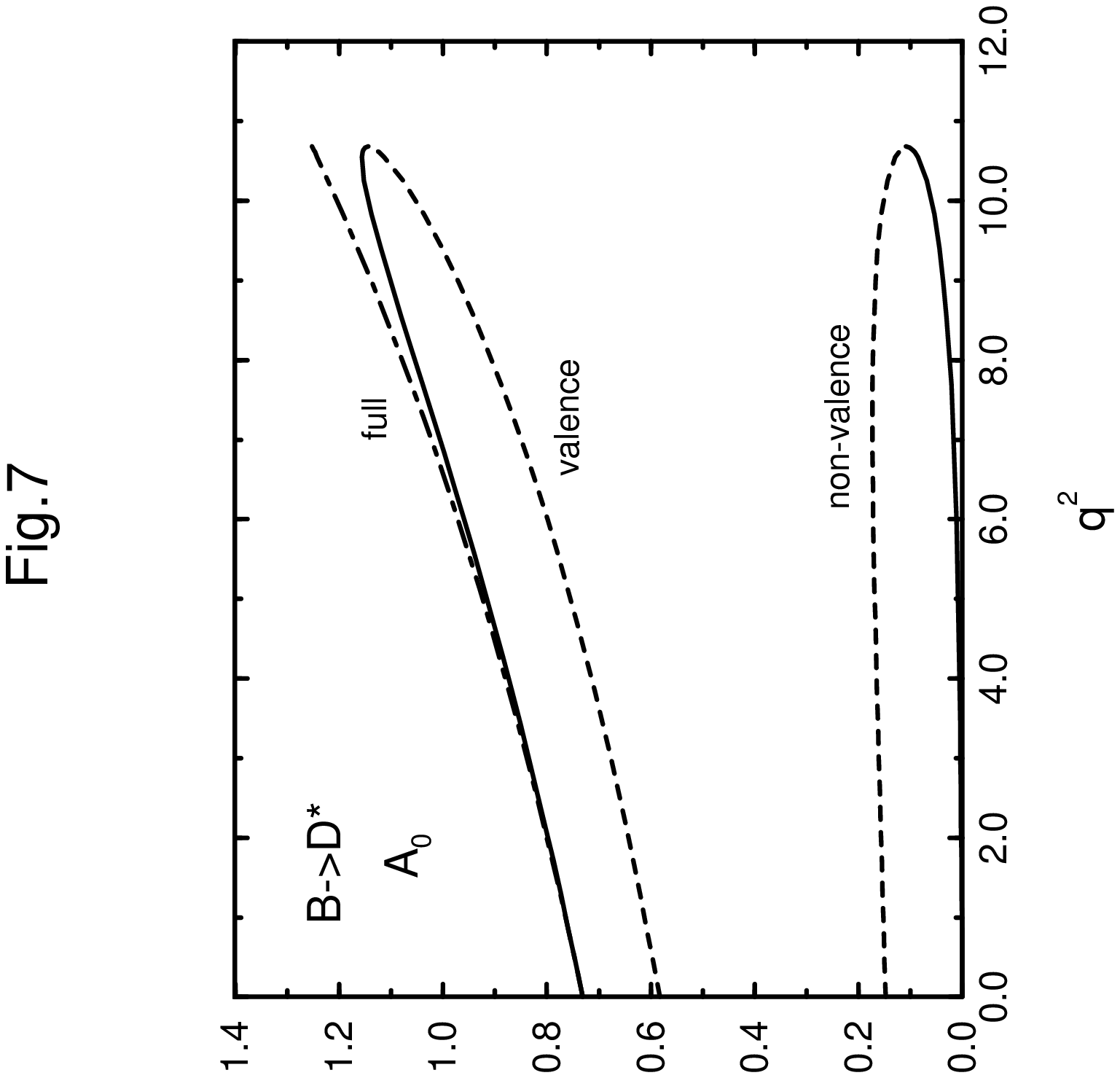}}
\end{figure}
\newpage

\begin{figure}[h]
\hskip 1cm
\hbox{\epsfxsize=16cm
      \epsfysize=20cm
      \epsffile{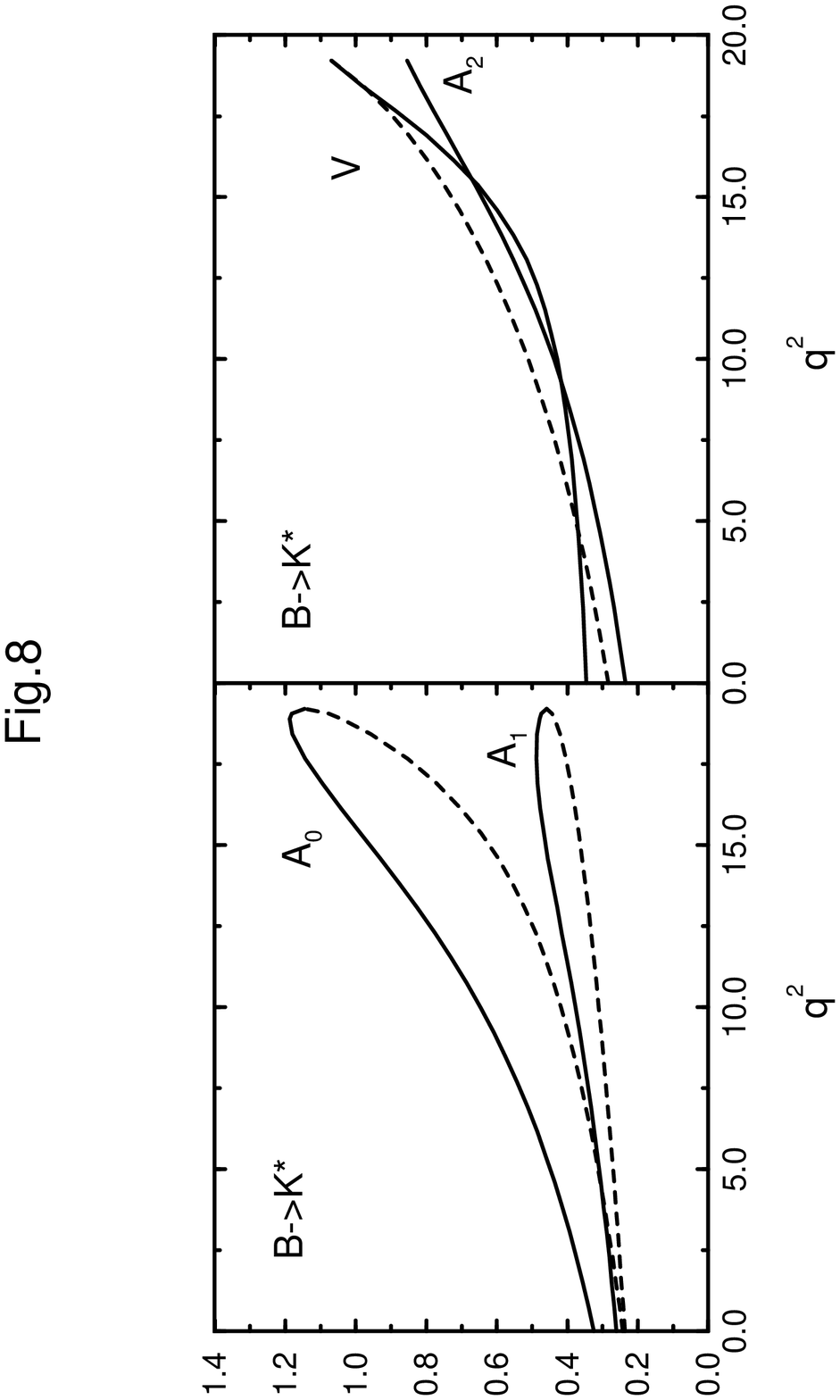}}
\end{figure}
\newpage

\begin{figure}[h]
\hskip 1cm
\hbox{\epsfxsize=16cm
      \epsfysize=20cm
      \epsffile{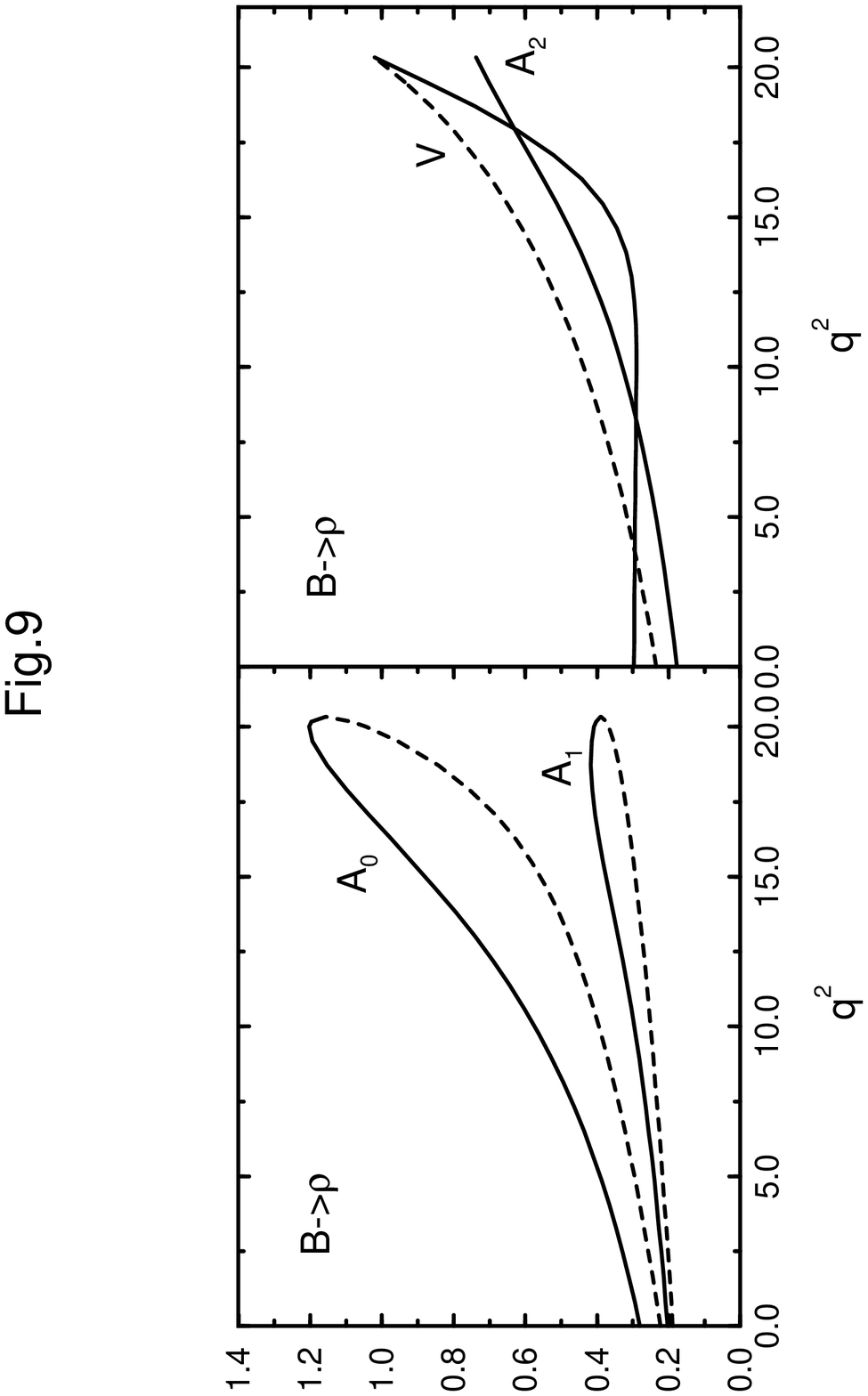}}
\end{figure}
\newpage

\begin{figure}[h]
\hskip 1cm
\hbox{\epsfxsize=16cm
      \epsfysize=20cm
      \epsffile{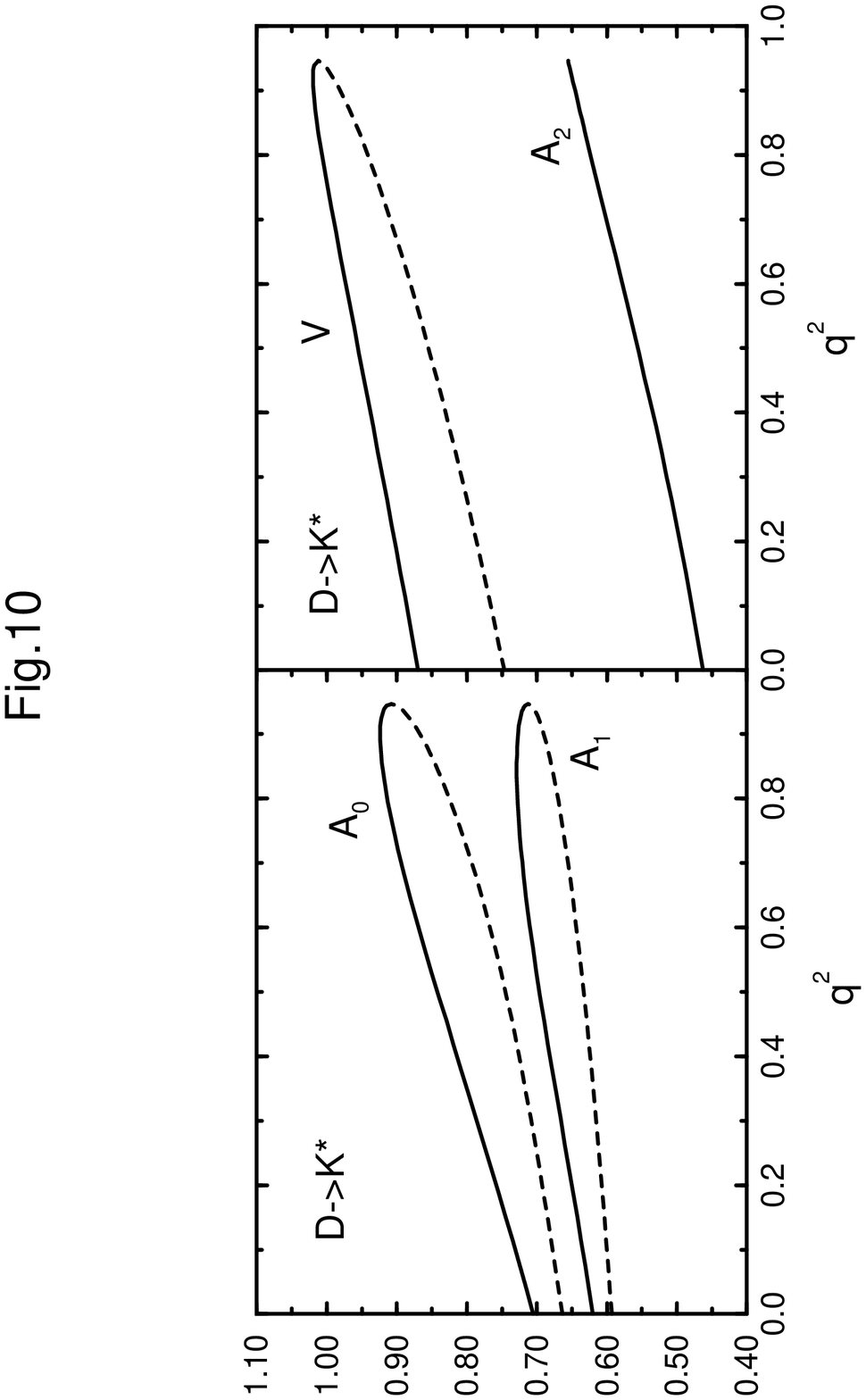}}
\end{figure}
\newpage

\begin{figure}[h]
\hskip 1cm
\hbox{\epsfxsize=16cm
      \epsfysize=20cm
      \epsffile{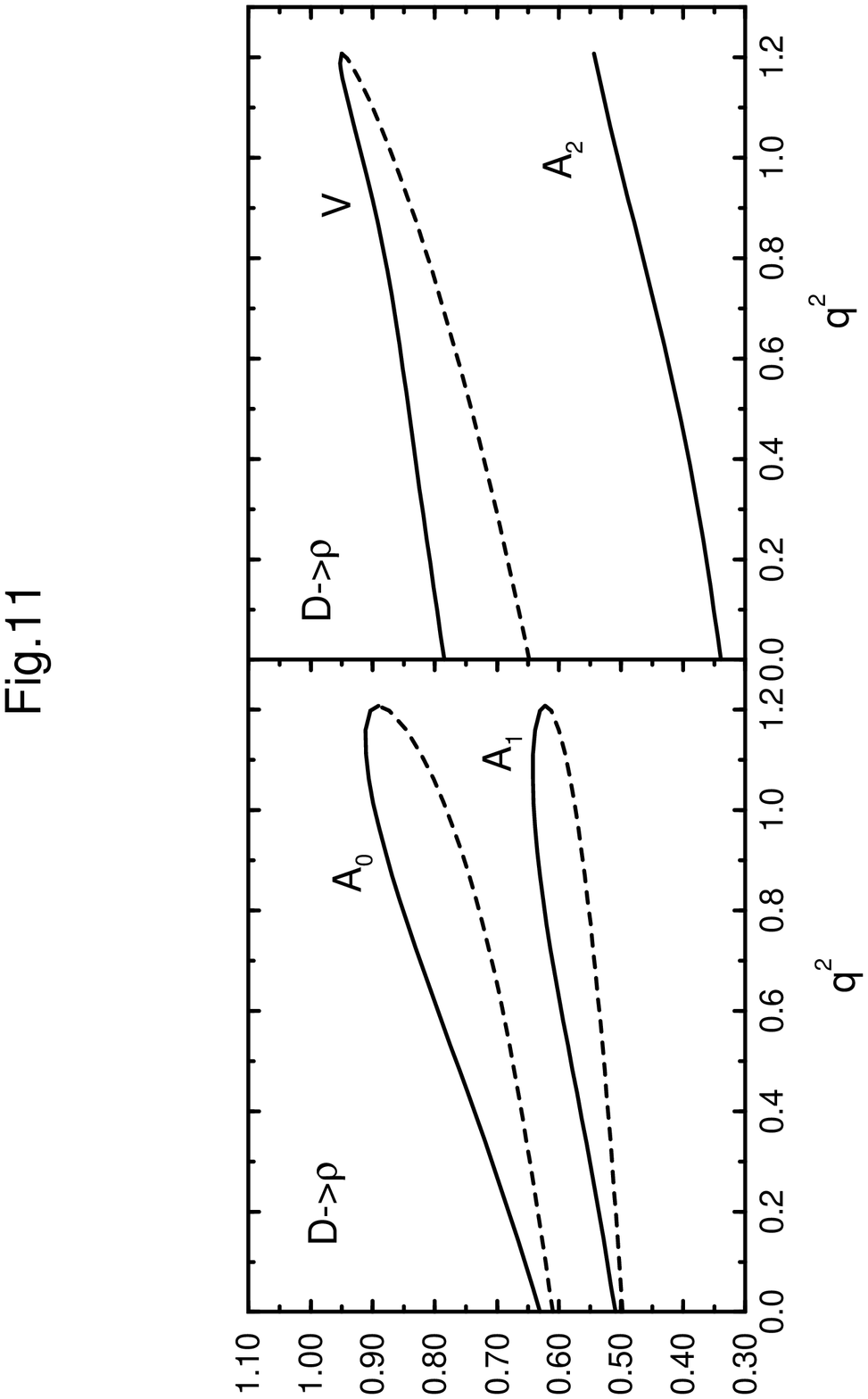}}
\end{figure}
\newpage

\end{document}